\documentclass[12pt,preprint]{aastex}
\bibpunct[; ]{(}{)}{;}{a}{}{;}

\shorttitle{The ACS Virgo Cluster Survey. VII. Dwarf-Globular Transition Objects}
\shortauthors{Ha\c{s}egan \etal}

\begin{document}


\title{The ACS Virgo Cluster Survey. VII. Resolving the Connection 
Between Globular Clusters and Ultra-Compact Dwarf Galaxies\altaffilmark{1}}


\author{Monica Ha\c{s}egan\altaffilmark{2,3},
Andr\'es Jord\'an\altaffilmark{2,4,5,6},
Patrick C\^ot\'e\altaffilmark{2,7},
S.G. Djorgovski\altaffilmark{8},
Dean E. McLaughlin\altaffilmark{9},
John P. Blakeslee\altaffilmark{10},
Simona Mei\altaffilmark{10},
Michael J. West\altaffilmark{11},
Eric W. Peng\altaffilmark{2,7},
Laura Ferrarese\altaffilmark{2,7},
Milo\v s Milosavljevi\'c\altaffilmark{8,12},
John L. Tonry\altaffilmark{13},
David Merritt\altaffilmark{14}}


\altaffiltext{1}{Based on observations with the NASA/ESA {\it Hubble
Space Telescope} obtained at the Space Telescope Science Institute,
which is operated by the association of Universities for Research in
Astronomy, Inc., under NASA contract NAS 5-26555.}
\altaffiltext{2}{Department of Physics and Astronomy, Rutgers, The
State University of New Jersey, Piscataway, NJ 08854;
mhasegan@physics.rutgers.edu}
\altaffiltext{3}{Institute for Space
Sciences, P.O.Box MG-23, RO 77125, Bucharest-Magurele, Romania}
\altaffiltext{4}{Claudio Anguita Fellow}
\altaffiltext{5}{Astrophysics, Denys Wilkinson Building, University of
Oxford, 1 Keble Road, Oxford, OX1 3RH, UK} 
\altaffiltext{6}{European
Southern Observatory, Karl-Schwarzschild-Str. 2, 85748 Garching,
Germany; ajordan@eso.org} 
\altaffiltext{7}{Herzberg Institute of
Astrophysics, National Research Council of Canada, 5071 West Saanich
Road, Victoria, BC, V9E 287, Canada; Patrick.Cote@nrc-cnrc.gc.ca,
Eric.Peng@nrc-cnrc.gc.ca, Laura.Ferrarese@nrc-cnrc.gc.ca}
\altaffiltext{8}{California Institute of Technology, Pasadena, CA
91125; george@astro.caltech.edu, milos@tapir.caltech.edu}
\altaffiltext{9}{Space Telescope Science Institute, 3700 San Martin
Drive, Baltimore, MD 21218; deanm@stsci.edu}
\altaffiltext{10}{Department of Physics and Astronomy, The Johns
Hopkins University, 3400 North Charles Street, Baltimore, MD
21218-2686; jpb@pha.jhu.edu, smei@pha.jhu.edu}
\altaffiltext{11}{Department of Physics and Astronomy, University of
Hawaii, Hilo, HI 96720; westm@hawaii.edu} 
\altaffiltext{12}{Sherman
M. Fairchild Fellow} 
\altaffiltext{13}{Institute for Astronomy,
University of Hawaii, 2680 Woodlawn Drive, Honolulu, HI 96822;
jt@ifa.hawaii.edu} 
\altaffiltext{14}{Department of Physics, Rochester
Institute of Technology, 84 Lomb Memorial Drive, Rochester, NY 14623;
merritt@mail.rit.edu}


\slugcomment{To appear in the Astrophysical Journal, 10 July 2005 issue}

\begin{abstract}
We investigate the connection between globular clusters and
ultra-compact dwarf galaxies (UCDs) by examining the properties of ten
compact, high-luminosity ($-11.8 \lesssim M_V \lesssim -10.8$) objects
associated with M87 (NGC~4486, VCC1316), the cD galaxy in the Virgo
Cluster.  These objects, most of which were previously classified as
M87 globular clusters, were selected from a combination of ground- and
space-based imaging surveys.  Our observational database for these
objects --- which we term DGTOs or ``dwarf-globular transition
objects" --- includes Advanced Camera for Survey (ACS) F475W and
F850LP imaging from ACS Virgo Cluster Survey, integrated-light
spectroscopy from Keck/ESI, and archival F606W WFPC2 imaging.  We also
present a search for DGTOs associated with other galaxies based on ACS
imaging for 100 early-type galaxies in Virgo.  Our main findings can
be summarized as follows:
\begin{itemize}

\item[(1)] Out of the six DGTOs in M87 with both ground-based
spectroscopy and {\it HST} imaging, we find two objects to have
half-light radii, velocity dispersions and mass-to-light ratios that
are consistent with the predictions of population synthesis models for
old, metal-rich, high-luminosity globular clusters.

\item[(2)] Three other DGTOs are much larger, with half-light radii
$r_h \sim$ 20~pc, and have $V$-band mass-to-light ratios in the range
$6 \lesssim {\cal M/L}_V \lesssim 9$.  These objects, which we
consider to be UCDs, resemble the nuclei of nucleated dwarf elliptical
galaxies in the Virgo Cluster, having similar mass-to-light ratios,
luminosities, and colors.

\item[(3)] The classification of the sixth object is more uncertain,
but it bears a strong resemblance to simulated ``stellar
superclusters" which are presumed to form through the amalgamation of
multiple young massive clusters.

\item[(4)] In general, the UCDs in M87 are found to follow the
extrapolated scaling relations of galaxies more closely than those of
globular clusters. There appears to be a transition between the two
types of stellar systems at a mass of $\approx 2\times10^6{\cal
M}_{\odot}$. We suggest that the presence of dark matter is the
fundamental property distinguishing globular clusters from UCDs.

\item[(5)] We identify a sample of 13 DGTO candidates from the
complete ACS Virgo Cluster Survey, selecting on the basis of
half-light radius, magnitude and color. For a number of these objects,
membership in Virgo can be established through radial velocities or
surface brightness fluctuation measurements with our ACS images. Three
of these DGTO candidates are embedded in low-surface brightness
envelopes.

\item[(6)] Five of the 13 DGTOs in Virgo are associated with a single
galaxy: M87.  This finding suggests that proximity to the Virgo center
may be of critical importance for the formation of these objects,
although we find M87 to be more abundant in DGTOs than would be
expected on the basis of its luminosity, the size of its globular
cluster system, or the local galaxy density.

\end{itemize}
These results show that distinguishing bonafide UCDs from
high-luminosity globular clusters requires a careful analysis of their
detailed structural and dynamical properties, particularly their
mass-to-light ratios.  In general, the properties of the UCDs in our
sample are consistent with models in which these objects form through
tidal stripping of nucleated dwarf galaxies.
\end{abstract}




\section{Introduction}
\label{sec:introduction}

A potentially new type of faint and compact stellar system has
recently been identified in the course of a spectroscopic survey in
the Fornax Cluster \citep{drink99,hilker99,drink00,phillipps01}.
Radial velocity measurements for these so-called ultra compact dwarf
galaxies (UCDs) reveal them to be Fornax members, and yet they are
unresolved in typical ground-based seeing. This observation places an
{\it upper limit} of $r_h \lesssim 50$~pc on the half-light radii of
these objects. This is considerably smaller than normal dwarf
galaxies, which have exponential scale-lengths $\approx$ 300~pc
\citep{deady02}, and at the same time is much larger than the
half-light radii of typical Galactic globular clusters, which have
median $r_h \approx 3$~pc.

The unusual properties of these objects have inspired a number of
different explanations for their origin. Possibilities include that
they are: (1) exceptionally large and luminous --- but otherwise
normal --- globular clusters which share a common origin with their
low-mass analogs; (2) the nuclei of nucleated dwarf elliptical (dE,N)
galaxies which happen to be embedded in envelopes of exceptionally low
surface-brightness; (3) the aggregate remains of massive young star
clusters which merge following their formation in gas-rich mergers
\citep{fellhauer02}; (4) the end-products of small-scale primordial
density fluctuations which collapsed in dense environments
\citep[see][]{phillipps01}; or (5) the tidally-stripped nuclei of
otherwise normal dE,N galaxies.  In fact, this latter scenario had
been considered several times in the past, beginning with attempts to
explain the overabundance of globular clusters associated with M87 in
terms of tidal stripping of dwarf galaxies. \citet{zinn88},
\citet{freeman90} and \citet{bassino94} had examined the possibility
that strong tidal effects on nucleated dwarf galaxies, in cluster
environments, might leave behind compact nuclei which resemble
globular clusters. More recent simulations
\citep[e.g.,][]{bekki01,bekki03,bekki04} have generally confirmed the
findings of \citet{bassino94} that, although non-nucleated dwarf
elliptical galaxies disintegrate completely after a few passages close
to a giant galaxy like M87, the nuclei of more massive dE,N galaxies
manage to survive long after the surrounding envelope has been tidally
stripped (a mechanism dubbed ``galaxy threshing'' by
\citealt{bekki01}).  \citet{phillipps01} and \citet{drink03} examine
the various possibilities, and argue that the compact objects in
Fornax are either the stripped remains of nucleated dwarfs, or a
genuinely new class of galaxy.

To date, the debate over the origin of UCDs has been guided largely by
the results of ground-based imaging and spectroscopy, which by
themselves provide little or no constraints on the internal structural
and dynamical properties of these objects. Such information is not
only essential for understanding their nature, but also for assessing
the extent to which they differ from massive, but otherwise ``normal",
globular clusters \citep[see also] []{mieske02}. However, {\it HST}
imaging has recently become available for several of these
objects. From an analysis of STIS images for five UCDs,
\citet{drink03} find typical absolute magnitudes and half-light radii
of $\langle{M_V}\rangle \sim -12$ and $\langle{r_h}\rangle \sim
16$~pc.  In light of the new $r_h$ measurements, rather than just
upper limits, the classification of UCDs as objects fundamentally
distinct from massive globular clusters seems less secure: for
instance, $\approx$ 9\% of the globular clusters in the Milky Way have
half-light radii of $10 \le r_h \le 22$~pc (the range spanned by the
objects in Fornax), although the majority of these globulars are
low-luminosity objects.  In any event, it is clear that reliable
information on the photometric, structural, dynamical properties for
an expanded sample of UCD candidates, preferably in new and different
environments, is needed to investigate the nature and homogeneity of
this potentially new type of galaxy. And improved observational
material for not just candidate UCDs, but also the exceptionally
bright globular clusters which are under-represented in Local Group
samples, is needed to investigate the extent to which these UCDs
differ from the most luminous globular clusters.

In this paper, we investigate the connection between globular clusters
and UCDs by studying objects selected from a combination of
ground-based catalogs of the M87 globular cluster system and {\it HST}
images acquired as part of the ACS Virgo Cluster Survey
\citep[hereafter Paper I]{cote04}.  To avoid biasing the discussion
and classification of these objects as either globular clusters or
UCDs, we henceforth refer to them as ``dwarf-globular transition
objects" or DGTOs. This term seems appropriate since these objects
have luminosities which place them simultaneously among the brightest
known globular clusters and the faintest dwarf elliptical galaxies.
Our goal is to combine our F475W and F850LP ACS imaging with archival
F606W WFPC2 imaging and ground-based spectroscopy from the Keck
telescope to identify and study these DGTOs.  In this paper, we focus
on the photometric, structural and dynamical properties of these
objects and we use this information to classify each DGTO as either a
globular cluster or a UCD; in a future paper, we shall investigate the
star formation and chemical enrichment histories of our program
objects using our Keck spectra (Ha\c{s}egan et~al. 2005; in
preparation).

The paper is organized as follows.  In \S\ref{sec:sample}, we discuss
the selection of the program objects. In \S\ref{sec:spectr}, we
describe the reduction of the spectroscopic data obtained with the
Echellette Spectrograph and Imager \citep[ESI;][]{sheinis02}.  The
measurement of radial velocities and internal velocity dispersions for
the program objects is presented in \S\ref{sec:disp}. The
determination of structural parameters for a subset of our program
objects, based on {\it HST} images, is discussed in
\S\ref{sec:photom}. We examine the scaling relations for these objects
in \S\ref{sec:scaling} and derive masses and mass-to-light ratios in
\S\ref{sec:mass}. Additional DGTO candidates identified in the ACS
Virgo Cluster Survey are reported in \S\ref{sec:candidates}. We
summarize our findings in \S\ref{sec:conclusions}.

\section{Sample Selection}
\label{sec:sample}

We begin by concentrating on ten DGTOs in the vicinity of VCC1316
(M87), the cD galaxy near the dynamical center of Virgo. All ten
objects have previously been classified as globular clusters
associated with M87, and all are confirmed members of Virgo, having
appeared in published photometric and radial velocity studies of the
M87 globular cluster system
\citep[e.g.,][]{strom81,cohen97,cohen98,hanes01,cote01}.  With
absolute magnitudes in the range $-11.8 \lesssim M_V \lesssim -10.8$,
these objects would be among the most luminous of M87's $\approx$
14,000 globular clusters \citep{mclaugh94}. In terms of luminosity
alone they have few, if any, counterparts in the Local Group.

As explained in the following sections, we have obtained ESI spectra
for all ten objects. {\it HST} imaging is available for six of the
ten. Among these six, three were observed with ACS in the F475W and
F850LP bandpasses as part of the ACS Virgo Cluster Survey, while
WFPC2/F606W images for three other objects are available from the {\it
HST} archive.

\subsection{Selection From Ground-Based Surveys}
\label{sec:ground}

Seven of our DGTOs (S314, S348, S417, S490, S804, S1370, S1538) were
selected from the $UBR$ photographic survey of the M87 globular
cluster system of \citet{strom81}.  In each case, membership in Virgo
was previously established through radial velocity measurements
\citep[e.g.,][]{mould90,cohen97,hanes01}.  These objects were selected
to be among the brightest of M87 globular clusters in the compilation
of \citet{hanes01}.

Figure 1 shows the positions of these seven objects in a $V$-band
image centered on M87 \citep{mclaugh94}. Note that they all fall
outside the field imaged by the ACS Virgo Cluster Survey, which is
indicated by the solid lines.

\subsection{Selection From the ACS Virgo Cluster Survey}
\label{sec:ACS}
 
Paper I gives a detailed description of the ACS Virgo Cluster survey
(GO-9401): a program to image, in two widely separated bandpasses
(F475W $\approx$ Sloan $g$ and F850LP $\approx$ Sloan $z$), 100
early-type members of the Virgo Cluster using the ACS instrument
\citep{ford98}.  Total exposure times for each galaxy are 750 sec in
F475W and 1210 sec in F850LP, respectively.  Given its depth and
uniformity, the survey offers a unique opportunity to study the
globular cluster systems of early-type galaxies in Virgo in a
systematic and comprehensive manner.  \citet[hereafter Paper
II]{jordan04a} discuss the pipeline developed for the reduction and
analysis of the 500 ACS Virgo Cluster Survey images: i.e., image
registration, drizzling strategies, the computation of weight images,
object detection, the identification of globular cluster candidates
and the measurement of their photometric and structural parameters.

Several DGTOs candidates were immediately apparent upon examining the
$\approx$ 2000 sources detected in our M87 field, the overwhelming
majority of which are globular clusters. Compared to the bulk of the
clusters in M87, these objects are unusually bright and spatially
extended.  Figure 2 shows a portion of the reduced ACS Virgo Cluster
Survey F475W image of M87, illustrating the position and appearance of
several DGTOs: the three objects for which we obtained ESI
spectroscopy (S928, S999, H8005) as well as two additional DGTOs
(H5065 and H8006; see \S8). The rectangle centered on each of the
first three objects shows the dimension and orientation of the
spectrograph slit used in the observations described in
\S\ref{sec:spectr}.
 
Table 1 contains coordinates, photometry and metallicities for the ten
DGTOs that are the focus of this paper.  The first column gives the
identification number of each object, where the names beginning with
an ``S'' are from \citet{strom81}, and those starting with an ``H''
are from \citet{hanes01}. The next four columns record the right
ascension and declination from \citet{hanes01}, the distance from the
nucleus of M87, and the apparent $V$ magnitude from
\citet{mclaugh94}. Columns 6--8 give the $g$ and $z$ magnitudes from
the ACS Virgo Cluster Survey, along with the dereddened $(g - z)_0$
color (assuming $A_g = 0.084$ and $A_z = 0.034$;
\citealt{schlegel98}). Column 9 gives the observed ($C-T_1$) color for
each object, taken from the database of \citet{hanes01}. The final
three columns record three different estimates of the metallicity for
each object: (1) that derived from the ($g - z$) colors and the
color-metallicity relation of \citet{bruzual03} for an assumed age of
13 Gyr; (2) that from the observed ($C-T_1$) color and the globular
cluster color-metallicity relation of \citet{cohen03}, assuming
$E(C-T_1)=2E(B-V$) \citep{geisler90} and $E(B-V)=0.023$
\citep{schlegel98}; and (3) that found from the Keck/LRIS spectroscopy
of \citet{cohen98}.  There is generally good agreement between the
various estimates, but since metallicities based on the ($C-T_1$)
index are available for all ten objects, we adopt these estimates in
the following analysis.

\section{ESI Spectroscopy}
\label{sec:spectr}

An instrument with both high spectral resolution and excellent
efficiency is needed to
measure the velocity dispersions of low-mass stellar systems, such as 
globular clusters or dwarf galaxies, at the distance of Virgo. 
Integrated-light spectra for the ten DGTOs
were obtained using the Keck 10m telescope and ESI on
29 April -- 1 May 2000 and 4 June 2003. During the first of these 
observing runs, we observed the seven objects which were selected from 
the ground-based studies; during
the second run, three objects identified from our ACS images 
were targeted. The same instrumental setup was used in both cases.

In echelle mode, ESI offers ten spectral orders, with complete wavelength
 coverage from 3900 {\AA} to 10900 {\AA} at a dispersion ranging from
0.15 {\AA} pixel$^{-1}$ (for $\lambda$ = 3900-4400 {\AA} in order \#15) to 
0.39 {\AA} pixel$^{-1}$ (for $\lambda$ = 9500-11000 {\AA} in order \#6). The
spectral dispersion, in units of velocity, is a nearly constant 
11.5 km~s$^{-1}$~pixel$^{-1}$. Each object was observed with a
0.75\arcsec$\times$20\arcsec slit, giving an instrumental velocity
resolution of $\approx$ 25 km~s$^{-1}$.

Tables 2 and 3 summarise the observation logs. In Table 2, we present
the exposure time, Heliocentric Julian Date, position angle of the ESI
slit, full width at half maximum for each object measured along the
slit's spatial direction, and the extraction half-width for each
observation (i.e. the distance along the slit, measured from the
object's center, over which the light profile was summed).  Objects in
our sample were observed once, with the exception of S999 and H8005,
which were observed twice.  Table 3 presents the right ascension,
declination, exposure time, Heliocentric Julian Date, spectral type,
visual magnitude and radial velocity for the radial velocity standard
stars observed during the two observing runs.

The processing of the raw data involved bias subtraction, finding and
tracing the apertures, flat normalization, cosmic ray removal, arc
extraction and spectral calibration. We used two independent paths for
the data reduction, based on the MAuna Kea Echelle Extraction (MAKEE)
\citep{barlow97} and IRAF reduction packages\footnote{IRAF is
distributed by the National Optical Astronomy Observatories, which are
operated by the Association of Universities for Research in Astronomy,
Inc., under cooperative agreement with the National Science
Foundation.}.  The velocity dispersion measurements described in the
following section were carried out using both the MAKEE and IRAF
processed spectra, yielding consistent results. Henceforth, the quoted
velocity dispersions refer to those measured using the MAKEE-processed
spectra.

Figure 3 shows the final spectra for the ten DGTOs in the spectral
range 5100-5300{\AA}. For S999 and H8005, we show the summed spectra
that were used in the analysis below.

\section{Velocity Dispersion Measurements}
\label{sec:disp}

For composite stellar systems, the observed spectrum is a
luminosity-weighted sum of individual stellar spectra shifted
according to their line-of-sight velocities. Assuming that the
individual spectra can be represented by a single template (generally
a reasonable assumption for globular clusters or early-type galaxies)
then the observed spectrum can be approximated as the convolution of
the template spectrum and the line-of-sight velocity distribution
(LOSVD), which acts as a broadening function. In the present case, our
goal is to estimate the internal velocity dispersion for each of our
DGTOs using the template (standard star) and observed spectra.

In order to assess the sensitivity of our results to the method used
to carry out the velocity dispersion measurements, three different
approaches were used: the Fourier Correlation Quotient method
\citep[FCQ;][]{bender90,bender94}, Penalized PiXel Fitting
\citep[pPXF;][]{capell04}, and the cross correlation method of
\citet{tonry79}.

The FCQ method constructs an estimate of the broadening function using
Fourier techniques, with the deconvolution based on the
template-object correlation function. While this algorithm is
relatively insensitive to template mismatch, it does require that the
absorption lines in the template spectrum be narrow compared to the
broadened lines of the object spectrum (i.e., that the object velocity
dispersion be large compared to the instrumental resolution;
\citealt{bender91}).  For our ESI observations, the instrumental
resolution ($\approx$ 25 km~s$^{-1}$) is comparable to that expected
for our targets. Indeed, this method failed to give stable results for
a few of our program objects. As a result, the FCQ results were not
used in the dynamical analysis presented below.

The pPXF algorithm considers the LOSVD of the stars as a Gauss-Hermite
series and attempts to recover it using a maximum penalized likelihood
formalism while working in pixel space \citep{merritt97,capell04}. It
has the advantage of being robust even when the data have low
signal-to-noise or when the observed LOSVD is not well sampled. Figure
4 gives an example of the pPXF output for S1538 and a template star
(HD154417) in the approximate wavelength range 5100--5300 {\AA}.

The method of \citet{tonry79} assumes that the LOSVD is a Gaussian
function with a dispersion equal to
\begin{equation}
\sigma = \sqrt{\mu^2 - 2\tau^2} \\
\label{eq1}
\end{equation}
where $\mu$ is the dispersion of the object-template cross-correlation
peak and the dispersion of the template auto-correlation peak is given
by $\tau$.  The Fourier auto- and cross-correlations were performed
with the RV.FXCOR task in IRAF. Measurements were carried out for
three spectral orders, spanning the wavelength range 4500--5800
{\AA}. Although the results obtained using different orders were
generally in good agreement, the radial velocities and velocity
dispersions used in the analysis below are those found using the
spectral region 5100--5300 {\AA} which, thanks to the high-S/N in this
region and the large number of sharp absorption lines including the
Mg~I triplet, produced the highest cross correlation peaks and the
best consistency among the various methods.

Tables 4 and 5 summarize our findings for the FXCOR and pPXF
methods. From left to right, these tables give the mean radial
velocity, the individual velocity dispersion measurements (from both
the FXCOR and pPXF methods and for a variety of templates), and the
average velocity dispersion from each method. It is clear that the two
methods give results that are in good agreement. The final column
gives the corrected central velocity dispersion, $\sigma_0$, obtained
using the pPXF measurements, after scaling upwards to account for the
blurring of the actual velocity dispersion profiles within the ESI
slit (see \S\ref{sec:photom}). Since an estimation of this correction
requires the intrinsic light profile to be known, we are able to give
central velocity dispersions only for those DGTOs with {\it HST}
imaging.  These six objects have $11 \lesssim \sigma_0 \lesssim
43$~km~s$^{-1}$, with mean $\langle\sigma_0\rangle \approx$ 28
km~s$^{-1}$. For comparison, Galactic globular clusters have central
dispersions $\sigma_0 \lesssim 18$~km~s$^{-1}$, with
$\langle\sigma_0\rangle \approx$ 7 km~s$^{-1}$ \citep{pryor93}.
Considering just the observed (i.e., uncorrected) dispersions, the
full sample of ten DGTOs has $9 \lesssim \sigma \lesssim
42$~km~s$^{-1}$, with mean $\langle\sigma\rangle \approx$ 28
km~s$^{-1}$.

\section{Structural Parameters}
\label{sec:photom}

\subsection{Objects with ACS Imaging}
 
As discussed in \S\ref{sec:ACS}, three DGTOs targeted for observation
with Keck/ESI were identified in the ACS Virgo Cluster Survey (S928,
S999 and H8005).  Since at the distance of Virgo, globular clusters
and even the most compact dE/dE,N galaxies are resolved in our ACS
images, it is possible to model directly the two-dimensional light
distribution of these objects. In fact, this is standard part of the
ACS Virgo Cluster Survey reduction pipeline, in which candidate
globular clusters are first identified on the basis of magnitude and
axis ratio (see Paper II). Since globular clusters in the Local Group
are nearly spherical systems, or at most modestly flattened, those
sources with axis ratios $\epsilon \equiv a/b \ge 2$ are discarded. In
addition, we discard sources brighter than approximately five
magnitudes above the expected turnover of the globular cluster
luminosity function at the distance of Virgo.  In principal, this last
criterion will exclude some extremely bright DGTO candidates, but the
adopted upper limits of $g \approx 19.1$ and $z \approx 18$ translates
into a luminosity cutoff $\sim$ 4$\times$ that of G1, one of the
brightest globular clusters of M31 \citep{meylan01}.

Photometric and structural parameters for all objects which satisfy
the above criteria were derived by fitting the two-dimensional ACS
surface brightness profiles (in both the F475W and F850LP filters)
with Point Spread Function (PSF)-convolved isotropic, single-mass
\citet{king66} models (see Paper II and \citealt[hereafter Paper
III]{jordan04b}). As described in Paper II, empirical PSFs in the
F475W and F850LP filters were derived using DAOPHOT II
\citep{stetson87,stetson93} and archival observations of fields in the
outskirts of the Galactic globular cluster NGC~104 (47 Tucanae).  The
King concentration index, $c$, half-light radius, $r_h$, and total
magnitude were derived for each globular cluster candidate, in both
filters. For King models, $r_h$ and the core radius, $r_c$, are
uniquely related for a given concentration index, so the fitted $r_h$
and $c$ were used to find $r_c$ for each object
\citep[see][]{mclaugh00}.

Observed and derived structural parameters for the three DGTOs
selected from the ACS Virgo Cluster Survey are presented in Table
6. The first two columns of this table record the object name and
ellipticity measured from the ACSVCS images using PSF-convolved King
models that allow the inclusion of ellipticity.  The next six columns
give the King concentration indices, half-light radii and core radii,
in both the $g$ and $z$ bands. The mean $V$-band surface brightness
within the half-light radii measured in the separate bandpasses,
\begin{equation}
\begin{array}{rrr}
\langle \mu_{V}^h \rangle_g & = & V - A_V + 0.7526 + 2.5\log{(\pi r_{h,g}^{2})} \\
\langle \mu_{V}^h \rangle_z & = & V - A_V + 0.7526 + 2.5\log{(\pi r_{h,z}^{2})} \\
\end{array}
\label{eq2}
\end{equation}
is given in the next two columns. These were calculated using the $V$
magnitudes presented in Table~1, assuming A$_V$ = 3.24E($B-V$) and a
color excess of E($B-V$) = 0.023 from \citet{schlegel98}.  The final
six columns report model-related parameters used in the computation of
King masses for our DGTOs, as explained in~\S\ref{sec:mass}. Note that
the half-light radii measured in the two bands show a small but
systematic difference of $\approx$ 7\% in the sense that the $z$-band
radii are larger, presumably a consequence of uncertainties in the
PSFs. In the analysis that follows, we use structural parameters
averaged between the two filters.

For the three DGTOs in our ACS field, Figure 5 shows radial profiles
in the F475W filter measured with IRAF task PLOT.PRADPROF.  The
adopted background value for each object is shown as the horizontal
dotted line; the mean F475W PSF for the M87 frame is shown as the
dashed profile. The best-fit, PSF-convolved King model is shown as the
solid curve (for two objects, S999 and S928, we notice a small $\sim
10- 20\%$ discrepancy in the central part of their profiles with
respect to the fit), the fitted parameters of which are given in Table
6. The arrow in each panel indicates the fitted half-light radius,
demonstrating the large spatial extent of these objects: $\langle
r_h\rangle = 0\farcs3$, or $\simeq$ 24~pc for our adopted M87 distance
of 16.1~Mpc \citep{tonry01}.

Figure~\ref{fig06} shows the distribution of half-light radii, in both
the $g$ and $z$ bandpasses, for the $\approx$ 2000 cataloged objects
in our M87 field.  We find median sizes of $r_{h,g}$ = 2.34~pc and
r$_{h,z}$ = 2.53~pc for these objects, with an $rms$ scatter of
$\approx$ 0.6~pc.  The vertical lines in Figure~\ref{fig06} show the
median half-light radius of Galactic globular clusters, $r_h$ =
3.2~pc.  Taken at face value, this result suggests that the globular
clusters in M87 are $\sim$ 30\% more compact than those in the Milky
Way. A more detailed discussion of the half-light radii and other
structural parameters for globular clusters in the ACS Virgo Cluster
Survey will be presented in a future
paper in this series.
 
The large squares in Figure~\ref{fig06} show the location of S999,
S928 and H8005 in the size-magnitude plane.  Not only are these DGTOs
among the brightest sources in the M87 field, they are far larger than
the vast majority of globular clusters belonging to this galaxy.  For
comparison, we show in Figure~\ref{fig06} the position of $\omega$
Centauri, G1 and M54, {\it as they would appear at the distance of
M87.} These objects, which are among the brightest of the $\sim$ 500
cataloged globular clusters in the Local Group, have sometimes been
identified as the nuclei of dwarf galaxies in various stages of
disruption and, hence, DGTOs in their own right
\citep{freeman93,meylan01,gnedin02,bekki-free03}.  In plotting these
objects in Figure~\ref{fig06}, we combine their absolute $V$
magnitudes and metallicities
\citep{harris96,meylan01,barmby02,mclaugh05} with the ($V-z$)-[Fe/H]
and ($g-z$)-[Fe/H] relations in Paper III to estimate their absolute
magnitudes in the $g$ and $z$ bandpasses.  Each object is then shifted
to our adopted distance of 16.1~Mpc for M87 and reddened
appropriately. We find $g$ $\approx$ 20.25 and $z$ $\approx$ 21.22 for
$\omega$ Cen, $g$ $\approx$ 20.62 and $z$ $\approx$ 19.46 for G1, and
$g$ $\approx$ 21.51 and $z$ $\approx$ 20.53 for M54. Half-light radii
are taken from \citet{harris96} and \citet{mclaugh05}.

It is clear from this figure that S999, S928 and H8005 are
objects quite unlike $\omega$ Cen, G1 and M54. They are, on average,
twice as luminous as the Local Group clusters, although this result
in itself is perhaps not surprising given that the sample of clusters
in our M87 field is 3--4$\times$ larger than is available in
the entire Local Group.  More significantly, the 
half-light radii of S999, S928 and H8005 are roughly five times
larger than those of the brightest Local Group clusters. Indeed, few
clusters of this size exist in the Local Group; from the 
catalog of $\sim$ 150 Galactic globular clusters given in 
\citet{harris96}, we find only a single object,
Pal~14, with $r_h > 20$~pc. But with an absolute magnitude of
$M_V \approx -4.7$, Pal~14 is hundreds of times fainter than
S999, S928 and H8005. In fact, the Galactic globular cluster which
most closely resembles the DGTOs in M87 is NGC~2419, which is
both large ($r_h \approx 20$~pc) and bright ($M_V = -9.3$). It is 
denoted in Figure~\ref{fig06} by the cross, where we have used
the procedure described above to estimate $g \approx 22.15$
and $z \approx 21.24$.

\subsection{Objects with WFPC2 Imaging}
\label{sec:wfpc2}
 
\begin{figure}
\plotone{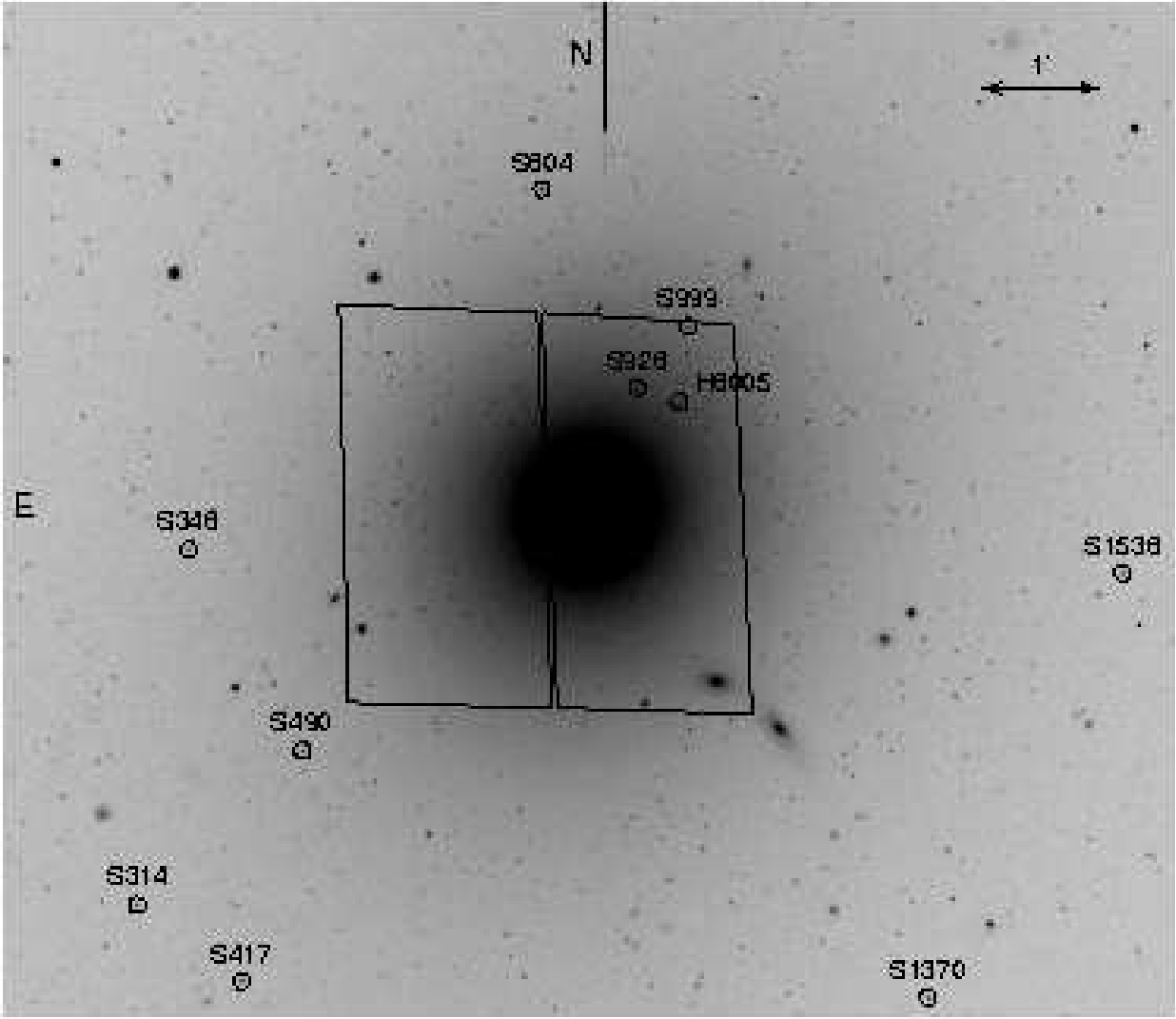}
\caption{$V$-band image of M87 obtained with the KPNO 4m telescope,
showing the location of our ten program objects (see Table~1). The
outlined regions show the footprint of ACS Wide Field Channel images
of M87 (VCC1316) from the ACS Virgo Cluster Survey.
\label{fig01}}
\end{figure}

Seven objects having ESI spectra fall outside the ACS field of view
(Figure~\ref{fig01}). A search of the {\it HST} archive revealed WFPC2
imaging in the F606W filter for three of these objects (S314, S427 and
S490), with exposure times between 500 and 1200 seconds.  These images
were retrieved from the archive and reduced using standard WFPC2
reduction procedures. In order to clean the images of cosmic rays, we
combined two images for each object using the WFPC.CRREJ task in IRAF,
taking care not to alter the cores of the objects. Structural
parameters for these objects were then derived using the same
procedure as described above. Parameters were derived using two
different PSF estimates: (1) a theoretical estimate for the F606W
filter calculated with Tiny Tim software \citep{krist95}; and (2) an
empirical estimate of the PSF for the F555W filter based on archival
observations (P.B. Stetson, private communication). Structural
parameters obtained using the two different techniques were in good
agreement, although the residuals were somewhat smaller in the latter
case. The results presented in Table 7 are those obtained with the
empirical F555W PSF. Two objects (S314 and S490) have $r_h \sim
3.4$~pc. This is close to the mean for globular clusters in the Milky
Way and just slightly larger than the average M87 globular
cluster. The third object (S417) is considerably larger, with $r_h
\approx 14$~pc. Clusters of this size are rare in the Milky Way, but
not entirely absent: for instance, among the 141 clusters with
measured half-light radii \citep{harris96}, ten objects (or 7\% of the
sample) have $r_h \gtrsim 14$~pc.  On the other hand, nine of these
ten objects are very faint, Palomar-type clusters, with $-6.7 \lesssim
M_V \lesssim -4.7$.  The tenth object is NGC~2419 which, as pointed
out above, happens to be one of the most luminous clusters in the
Milky Way (yet still a full order of magnitude fainter than S417).

For all DGTOs with {\it HST} imaging, we have calculated central
velocity dispersions, $\sigma_0$, by correcting the observed values of
$\langle\sigma\rangle$ to account for the blurring of the light
profile within the spectrograph.  We used the best-fit King model
parameters obtained from our ACS and WFPC2 imaging, and convolved the
corresponding King model surface brightness and velocity dispersion
profiles with various Gaussian kernels to approximate the effects of
ground-based seeing. The adopted kernel was taken to be that which
produced a FWHM for each object which matched the FWHM measured along
the ESI slit (see Table~2). We then determined the correction factor
for each object needed to convert the measured dispersion to the true,
central value. These correction factors were always $\lesssim
15$\%. The resulting values of $\sigma_0$ are tabulated in the final
columns of Tables 4 and 5.

\section{Scaling Relations}
\label{sec:scaling}

Correlations between the physical properties of stellar systems
hold important clues for understanding their formation and evolution
\citep[e.g.,][]{korm85,djor93}. Numerous studies of globular cluster
scaling relations have shown them to be comparatively ``simple"
systems: i.e., well approximated by isotropic, equilibrium King models
with a constant mass-to-light ratio of $1.45\pm0.1$ (see
\citealt{mclaugh00} and references therein).
By constrast, studies of other low-mass stellar systems, such as
dE and dSph galaxies, indicate that dark matter or non-equilibrium
dynamics must be important. For instance, Local Group 
dSph galaxies have $3 \lesssim {\cal M/L_V} \lesssim 84$ \citep{mateo98}
while intermediate values, $3 \lesssim {\cal M/L_V} \lesssim 6$, have been
measured for a handful of dE,N galaxies, and their nuclei, in the
Virgo Cluster \citep{geha02,geha03}. Meanwhile, \citet{drink03}
 have reported central velocity dispersions of 24--37 km~s$^{-1}$ and
$V$-band mass-to-light ratios in the range 2--4 for five faint compact
objects in the Fornax Cluster, and argued on this basis that they
constitute a fundamentally different class of object from globular
clusters.  In this section, we investigate the scaling relations
of our DGTOs and present a comparison with those of
globular clusters and galaxies.

In the scaling relations shown in Figures~\ref{fig07}--\ref{fig11}, we
plot the location of globular clusters belonging to the Galaxy
\citep{mclaugh05}, M31 (using the data collated in
\citealt{mclaugh05}, in preparation) and NGC~5128
\citep{martin04,harris02}; Local Group dSph galaxies \citep{mateo98};
Virgo dE,N galaxies and their nuclei \citep{geha02,geha03}; the UCDs
in Fornax \citep{drink03}; and the compact Local Group elliptical
galaxy M32 \citep{mateo98,marel98,graham02}. Our DGTOs (with data from
Tables~1, 4, 5, 7 and 8) are indicated by the large squares. The Local
Group globular clusters $\omega$~Cen, G1 and M54 are indicated by the
large filled triangles, while NGC~2419 is shown as the cross.

Figure~\ref{fig07} shows a particularly important scaling relation for
hot stellar systems: central velocity dispersion versus absolute
visual luminosity.  The behavior of high-luminosity galaxies is
indicated by the crosses, which show the bright ellipticals from
\citet{faber89}. The dashed line represents the Faber-Jackson (1976)
relation obtained by fitting to the bright ellipticals $({\cal L_V} >
10^{10} {\cal L_{\odot}}$) and finding the slopes that minimize the
$rms$ residuals perpendicular to the best-fit line, to allow for error
in both coordinates:
\begin{equation}
\begin{array}{rcl}
{\log \sigma_0} & = &  -0.347 + 0.25~{\log {\cal L_V}}. \\
\end{array}
\label{eq3}
\end{equation}
For comparison, the solid line shows the best-fit relation for 
Galactic globular clusters from \citet{mclaugh05}:
\begin{equation}
\begin{array}{rcl}
{\log \sigma_0} & = &  -1.781 + 0.50~{\log {\cal L_V}}. \\
\end{array}
\label{eq4}
\end{equation}
This relation also provides an adequate description of the globular
clusters in M31 and NGC~5128, as well as the nuclei of dE,N galaxies
in Virgo.  With $M_V \approx -10.8$ to $-11.8$, most of our program
objects lie close to the intersection of these two scaling relations;
certainly one cannot claim on the basis of this figure alone that the
DGTOs are inconsistent with the globular cluster relation. We believe,
that with the possible exception of UCD3 (the brightest object in the
study of \citealt{drink03}), this conclusion also applies to the compact
objects in Fornax. \citet{drink03} reported that they fall along the
extrapolation of the Faber-Jackson relation; however, compared to
Equation~\ref{eq3}, the galaxy scaling relation shown in their
Figure~3 overpredicts by $\sim$ 25\% the central velocity dispersion
at fixed luminosity.  Figure~\ref{fig07} suggests that discriminating
between bright globular clusters and bonafide UCDs based on their
location in the $M_V$-$\sigma_0$ plane is difficult at
best. Additional information (e.g., structural parameters and internal
dynamical properties) is needed to distinguish globular clusters from
UCDs in this luminosity range.

In Figure~\ref{fig08}, we combine our $\sigma_0$, $r_h$ and $M_V$
measurements to show the Virial Theorem for low-mass stellar systems:
${\cal M} \propto r_h{\sigma_0}^2$.  The lower dashed line in this
panel shows the expected relation for a constant mass-to-light ratios
of ${\cal M/L_V} = 1.45$, the value appropriate for non-core-collapsed
globular clusters in the Milky Way \citep{mclaugh00}. The fact that
the DGTOs lie above this relation suggests that this value may be
inappropriate for these objects. Indeed, this relation appears to
underpredict the mass-to-light ratios of systems brighter than $M_V
\approx -10.5$. Anticipating the results of \S7, the upper dashed line
shows the scaling relation for ${\cal M/L_V} = 5$, a relation which
better describes the brighter DGTOs and provides a superior
representation of the central nuclei in Virgo dE,N galaxies
\citep{geha02}.  The luminous globular clusters in NGC5128
\citep{martin04} and the majority of the Fornax UCDs \citep{drink03} 
appear to have intermediate mass-to-light ratios.

As is well known, elliptical galaxies fall along a two-dimensional
surface in the parameter space defined by size, surface brightness and
velocity dispersion (see \citealt{djor94} for a review):
\begin{equation}
\begin{array}{rcl}
\alpha\langle{\mu_V^h}\rangle & = & {\log \sigma_0} + \beta{\log
{r_h}} + \gamma.\\
\end{array}
\label{eq5}
\end{equation}
Analogs of this Fundamental Plane for globular clusters have been
presented by \citet{djor95}, and more recently, by
\citet{mclaugh00}. Figure~\ref{fig09} shows these two representations
of the globular cluster Fundamental Plane (upper and lower panels,
respectively). The dashed line in the upper panel shows a
least-squares fit for the Galactic globular clusters to the bivariate
correlation, with $\beta = -0.7$ from \citet{djor95}:
\begin{equation}
\begin{array}{rcl}
\alpha & = & -0.24\pm0.02 \\
\gamma & = & -4.83\pm0.26. \\
\end{array}
\label{figfp1}
\end{equation}
In the lower panel, the dashed line indicates the expected relation if
globular clusters obey a relation between binding energy and total
luminosity, $E_b \propto L^{2.05}$, which gives $\beta = -0.775$ and
\begin{equation}
\begin{array}{rcl}
\alpha & = & -0.205 \\
\gamma & = & -4.147. \\
\end{array}
\label{figfp2}
\end{equation}
This relation, from \citet{mclaugh00}, accurately describes the
globular clusters in the Milky Way \citep{djor95}, M31
\citep{barmby02}, NGC~5128 \citep{harris02} and M33
\citep{larsen02}. It also provides an adequate description of the six
objects from our survey, although there does appear to be a zeropoint
offset of $\approx$ 0.2, in the sense that the DGTOs lie slightly
above the dashed line. As discussed in Appendix A1 of
\citet{mclaugh00}, the zeropoint of the line defined by
Equation~\ref{figfp2} depends on mass-to-light ratio, with
$\Delta\gamma \approx 0.5\Delta\log{{\cal M/L_V}}$.  Thus, the
observed tendency for the program objects to lie above the cluster
relation --- drawn here at a constant a mass-to-light ratio of 1.45
--- suggests that some of the DGTOs have mass-to-light ratios which
are $\approx$ 2-3 $\times$ higher than globular clusters in the Milky
Way.  This same conclusion applies to the dE,N nuclei of
\citet{geha02}.  As before, the majority of the UCDs in Fornax 
show a more modest enhancement in mass-to-light ratio relative to
globular clusters in the Milky Way.  This findings are consistent with
conclusions based on Figure~\ref{fig07}.  Apart from the small
zeropoint difference, the DGTOs show good agreement with the
\citet{mclaugh00} relation so we conclude that, as a class, they obey
the Fundamental Plane defined by globular clusters.

\section{Masses and Mass-to-Light Ratios}
\label{sec:mass}

The mass-to-light ratio of a stellar system is important for
understanding its nature and origin, offering insight into the
underlying stellar populations, dark matter content, and even the
extent to which the system is in virial equilibrium.  For the six
objects having measured velocity dispersions and structural
parameters, we may derive masses using the King model approximation.
The King mass, ${\cal M}_{k}$, is given by
\begin{equation}
{\cal M}_{k}=\frac{9}{2 \, \pi \, G} \frac{\nu \, r_c \, \sigma_0^2}{\alpha \, p}
\label{eqking}
\end{equation}
\citep[e.g.,][]{dubath97} where $G$ is the gravitational constant,
$\sigma_0$ is central velocity dispersion and $r_c$ is the core
radius.  The parameters $\nu$, $p$ and $\alpha$ depend on the fitted
King model, and were calculated through spline interpolation of the
relations given in \citet{king66} and \citet{peter75}.  For $r_c$,
$\nu$, $p$ and $\alpha$, we take the average of the measurements in
$g$ and $z$.  Structural parameters used in the mass determinations
are presented in Table~7.

Our results are presented in Table 8, which records object name,
$V$-band absolute magnitude and luminosity, mass and mass-to-light
ratio. It is apparent that the mass-to-light ratios for our six DGTOs
are higher (i.e., $3 \lesssim {\cal M}_k/{\cal L_V} \lesssim 9$) than
the mean value of 1.45$\pm$0.1 for Galactic globular clusters, which
is again consistent with expectations based on the bivariate
correlations discussed in \S6.  We now turn to the question of whether
these mass-to-light ratios {\it require} these DGTOs to have a
fundamentally distinct nature from globular clusters.  We begin by
examining the scaling relations, as a function of dynamical mass, for
low- and intermediate-mass stellar systems.

For {\it bright} elliptical galaxies, studies of the Fundamental Plane
indicate a dependence between luminosity and mass-to-light
ratio. Using the dynamical masses for elliptical galaxies given by
\citet{marel91}, we find
\begin{equation}
\begin{array}{rcl}
{\cal M/L_V} & = & 6.3 \big( { {\cal L_V} / 10^{11} } \big )^{0.3}$$ \\
\end{array}
\label{eq9}
\end{equation}
where we have adopted $H_0 = 70$~km~s$^{-1}$~Mpc$^{-1}$ and
transformed his $R$ magnitudes into the $V$ band. Equation~\ref{eq9}
is equivalent to
\begin{equation}
\begin{array}{rcl}
{\log {\cal L_V}} & = &  1.924 + 0.769~{\log {\cal M}}. \\
\end{array}
\label{eq10}
\end{equation}
Combining this result with Equation~\ref{eq3} yields the relationship
between galaxy mass and central velocity dispersion given
below. Likewise, a fit to the bright ellipticals of \citet{faber89}
gives
\begin{equation}
\begin{array}{rcl}
{\log r_h} & = &  -4.806 + 0.80~{\log {\cal L_V}}. \\
\end{array}
\label{eq11}
\end{equation}
for the galaxy size-luminosity relation. With $r_h$ and $\cal M$
known, one can calculate the mean mass density within the half-light
(or half-mass) radius, $\langle\Sigma_h\rangle$, and its dependence on
mass. Combining Equations~\ref{eq3}, \ref{eq10} and \ref{eq11}, one
finds the following scaling relations for elliptical galaxies:
\begin{equation}
\begin{array}{rcl}
{\log \sigma}_{0} & = & 0.134 + 0.192~{\log {\cal M}} \\
{\log r}_{h} & = & -3.267 + 0.615~{\log {\cal M}}  \\
{\log \langle\Sigma}_{h}\rangle & = & 5.736 - 0.230~{\log {\cal M}}. \\
\end{array}
\label{eq12}
\end{equation}
These relations, extrapolated into the low- and intermediate-mass
regimes, are shown as the dashed lines in Figure~\ref{fig10}. From top
to bottom, this figure plots $\sigma_0$, $r_h$ and
$\langle\Sigma_h\rangle$ against ${\cal M}$ for galaxies in the range
$10^3-10^8 {\cal M}_{\odot}$.

For comparison, the solid lines in each panel show the best-fit
relations for the Galactic globular cluster system:
\begin{equation}
\begin{array}{rcl}
{\log \sigma}_{0} & = & -1.904 + 0.509~{\log {\cal M}} \\
{\log r}_{h} & = & 0.509~~{\rm (median)}\\
{\log \langle\Sigma}_{h}\rangle & = & -1.815 + {\log {\cal M}} \\
\end{array}
\label{eq13}
\end{equation}
where the second relation is the familiar result that globular
clusters sizes are independent of luminosity. Observational data for
the low- and intermediate-mass stellar systems shown in
Figures~\ref{fig07}-\ref{fig09} are plotted in Figure~\ref{fig10} with
the same symbols. Note that, in the case of the Fornax UCDs, masses
are not available on an individual basis, so we combine their visual
magnitudes with the mean mass-to-light ratio of ${\cal M/L}_V = 3$
reported by \citet{drink03} to estimate their
masses. Figure~\ref{fig10} also shows the predicted properties of UCDs
from two sets of numerical simulations: (1) those of \citet{bekki04},
who explored the possibility that UCDs form through mergers of
globular clusters within dwarf galaxies which then undergo tidal
stripping; and (2) those of \citet{fellhauer02}, who modeled UCD
formation as the amalgamation of young massive star clusters which
form in gas-rich mergers.  While these simulations are fully
independent, in both cases the formation of UCDs is presumed to
involve multiple mergers of star clusters.

Perhaps the most noteworthy feature of Figure~\ref{fig10} is the
apparent presence of two separate families in this low-mass regime,
with a transition at a few million solar masses; formally, the
globular cluster and galaxy scaling relations cross at ${\cal M}_t
\approx 2\times10^6{\cal M_{\odot}}$. This transition, which is
reminiscent of that observed at $M_V \approx -10.5$ in the
Faber-Jackson and Virial Theorem scaling relations presented in
Figures~\ref{fig07} and \ref{fig08}, takes the form of a change in
slope, or a ``break", in the $\log{\sigma_0}-\log{\cal M}$ relation
shown in the upper panel. That this truly represents a fundamental
transition between two families of stellar systems is best seen in the
middle panel of this figure: above ${\cal M}_t$, the size of an object
increases in proportion to its mass, while below ${\cal M}_t$,
globular clusters have a characteristic size of $r_h \sim 3$~pc
irrespective of mass. As a consequence, the $\log{\Sigma_h}-\log{\cal
M}$ relation shown in the lower panel shows a peak near ${\cal
M}_t$. In other words, for globular clusters, ${\Sigma_h}$ increases
with mass, while the converse is true of galaxies.

Focusing on the six DGTOs shown in Figure~\ref{fig10}, we see that two
objects (S314 and S490) lie on the extrapolation of the
$\log{\sigma_0}-\log{\cal M}$ relation for globular clusters {\it and}
have sizes of $r_h \approx 3$~pc. As a result, they also lie along the
extrapolation of the globular cluster $\log{\Sigma_h}-\log{\cal M}$
relation shown in the lower panel.  On this basis, these two objects
should probably be considered globular clusters, albeit ones of
unusually high mass and luminosity.

Three additional objects (S417, S928 and S999) fall along the galactic
scaling relations in all three panels. Although the distinction
between the globular cluster and galaxy scaling relations is modest in
terms of velocity dispersion, there is a clear separation in terms of
$r_h$ and $\Sigma_h$.  In general, these objects occupy roughly the
same positions in the $\log{\sigma_0}-\log{\cal M}$, $r_h-\log{\cal
M}$ and $\log{\Sigma_h}-\log{\cal M}$ planes as the simulated UCDs of
\citet{bekki03}. As such, they must be considered prime UCD
candidates.

The sixth and final DGTO, H8005, presents more of a puzzle. It falls
below both the globular cluster and galaxy relations in the
$\log{\sigma_0}-\log{\cal M}$ plane and, with $r_h \sim 29$~pc is
larger than one would expect for a globular cluster, or a dwarf
galaxy, at a mass of $\approx 5.5\times10^6{\cal M_{\odot}}$.
Moreover, it is an extreme outlier in the $\log{\Sigma_h}-\log{\cal
M}$ plane, with a mean half-mass density that is roughly an order of
magnitude lower than any other object in our sample. Interestingly,
this object bears a striking resemblance to the E01 simulation of
\citet{fellhauer02} at an age of 10 Gyr. If this agreement is not
merely coincidental, then H8005 may represent the remains of a
``stellar supercluster" which formed through the multiple mergers of
young massive clusters in the distant past.  Based on the available
evidence, we provisionally classify H8005 as a UCD, but note that its
structural properties may be fundamentally different from those of
S417, S928 and S999.

Taken together, the scaling relations in Figure~\ref{fig10} offer hope
for distinguishing globular clusters from UCDs in this low-mass
regime. As an additional aid in classifying DGTOs, we now turn our
attention to their measured mass-to-light ratios.

At fixed age, the mass-to-light ratio of a stellar system depends on
its metal abundance, with ${\cal M/L_V}$ increasing towards higher
[Fe/H].  Mass-to-light ratios and metallicities are available for six
DGTOs in our sample (see Tables~1 and 8). Figure~\ref{fig11} plots
mass-to-light ratio versus metallicity for these objects, along with
those of globular clusters in the Milky Way and NGC~5128 (with
$\omega$ Cen, G1 and M54 indicated by the large triangles), and Local
Group dwarf spheroidal galaxies.  The large open triangle in this
figure shows the {\it mean} location of the Fornax UCDs, which have
$\langle{\cal M/L_V}\rangle = 3\pm1$ \citep{drink03}.  We have
estimated crude metallicities for these objects by combining the mean
colors of $\langle(B-V)\rangle = 0.89$ and $\langle(V-I)\rangle =
1.09$ reported by \citet{karick03} with the respective
color-metallicity relations of \citet{couture90} and
\citet{kissler98}. This yields [Fe/H]$_{BV} \sim -0.4$ and
[Fe/H]$_{VI} \sim -0.9$; we consequently adopt $\langle$[Fe/H]$\rangle
\sim -0.65\pm0.35$ for the Fornax UCDs.

The five curves show theoretical predictions for the dependence of
mass-to-light ratio on metallicity, according to the population
synthesis models of \citet{bruzual03} with the disk-star initial mass
function of \citet{chabrier03}.  Curves show the expected behavior for
single-burst populations with ages of 7, 9, 11, 13 and 15 Gyr.

In accordance with the preceding discussion, all six DGTOs are seen to
have mass-to-light ratios larger than the average value of 1.45 for
Galactic globular clusters. At first glance, this might appear
surprising since our sample includes two objects, S314 and S490, which
we have argued above are probably globular clusters. However, the
higher mass-to-light ratios of these objects (${\cal M/L_V} \approx$
3--4) are perfectly consistent with the model predictions since these
are also the two most metal-rich objects in our small
sample. Similarly, the mass-to-light ratios of ${\cal M/L_V}$ = 2--4
reported by \citet{drink03} for the compact Fornax objects are also
consistent with model predictions for old globular clusters with
[Fe/H] $\gtrsim -1$.  In other words, mass-to-light ratios alone
provide no justification for the classification of these objects as
UCDs.

On the other hand, three objects (S417, S928 and S999) are found to
have mass-to-light ratios in the range $6 \lesssim {\cal M/L_V}
\lesssim 9$.  Mass-to-light ratio this large are much more difficult
to explain in terms of age and metallicity effects.  Indeed, for no
age/metallicity combination is it possible to produce mass-to-light
ratios as large as $\sim 9$ (as is the case for S999) without
resorting to extreme initial mass functions (see below). Whether this
indicates the presence of dark matter in these systems, or that the
fundamental assumption of virial equilibrium underlying
Equation~\ref{eqking} is invalid, remains to be seen. At the very
least, it is clear that their mass-to-light ratios set these objects
apart from normal globular clusters. It is also worth noting that
their mass-to-light ratios appear to more closely resemble those
measured for Virgo dE,N galaxies (with a median value in the $V$ band
of 5.5; \citealt{geha02}) than for globular clusters.

Finally, the measured mass-to-light ratio for H8005, ${\cal M/L_V}
\approx 3$, is slightly larger that predicted by the population
synthesis models for a globular cluster at this metallicity although
the discrepancy is not statistically significant given the measurement
errors.  This observation is again consistent with the stellar
supercluster scenario of \citet{fellhauer02} who note that, after
$\approx$ 10 Gyr of stellar fading with no morphological evolution,
their merged superclusters should have mass-to-light ratios identical
to those of normal star clusters at this age (and metallicity).

All in all, Figure~\ref{fig11} reinforces the DGTO classifications
presented above. To summarize, we divide our sample of six DGTOs into
three categories:
\begin{itemize}
\item[1.] Two {\it probable} globular clusters: S314 and S490 
\item[2.] Three {\it probable} UCDs: S417, S928 and S999
\item[3.] One {\it possible} UCD: H8005.
\end{itemize}

The question remains, however, as to {\it why} the mass-to-light
ratios of the probable UCDs resemble those of dwarf nuclei and, indeed
how such mass-to-light ratios might arise in the first place.  In the
galaxy threshing scenario, the mass-to-light ratios of UCDs should
mimic those of the dE,N nuclei, which does indeed seem to be the
case. On the other hand, the dE,N nuclei are taken to be baryon
dominated at all times in the simulations of \citet{bekki01}. 
Simply stated, a physical explanation for the high
mass-to-light ratios observed for the most probable UCDs is lacking,
although some possibilities exist (e.g., the truncation of the initial
mass function during starbursts; \citealt{charlot93}). We shall return
to the issue of dE,N nuclei in a future paper in this series.

\section{A Search for Additional DGTOs in the ACS Virgo Cluster Survey}
\label{sec:candidates}

Obviously, it is of interest to know if there are additional DGTOs in
the ACS Virgo Cluster Survey, either in M87 or belonging to other
program galaxies. We have searched our object catalogs for the
complete sample of 100 galaxies, selecting candidate DGTOs on the
basis of magnitude, half-light radius and color. Guided by the
location of the DGTOs in Figure~\ref{fig06}, we adopt
\begin{equation}
\begin{array}{rrcrr}
18 & \le & g & \le & 21 \\
17 & \le & z & \le & 20 \\
10 & \le & {r_h} {\rm (pc)} & \le & 100 \\
\end{array}
\label{sel1}
\end{equation}
where the condition on $r_h$ is required in both bands. 
In addition, we restrict the dereddened color to the range
\begin{equation}
\begin{array}{rrcrr}
0.5 & \le & (g-z)_0 \le & 1.6 \\
\end{array}
\label{sel2}
\end{equation}
which is appropriate for globular clusters and dwarf galaxies having
normal stellar populations (see Figures~5 and 6 of Paper I). At the
same time, this color selection minimizes the number of high-redshift
elliptical galaxies which fall into the sample.

\begin{figure}
\plotone{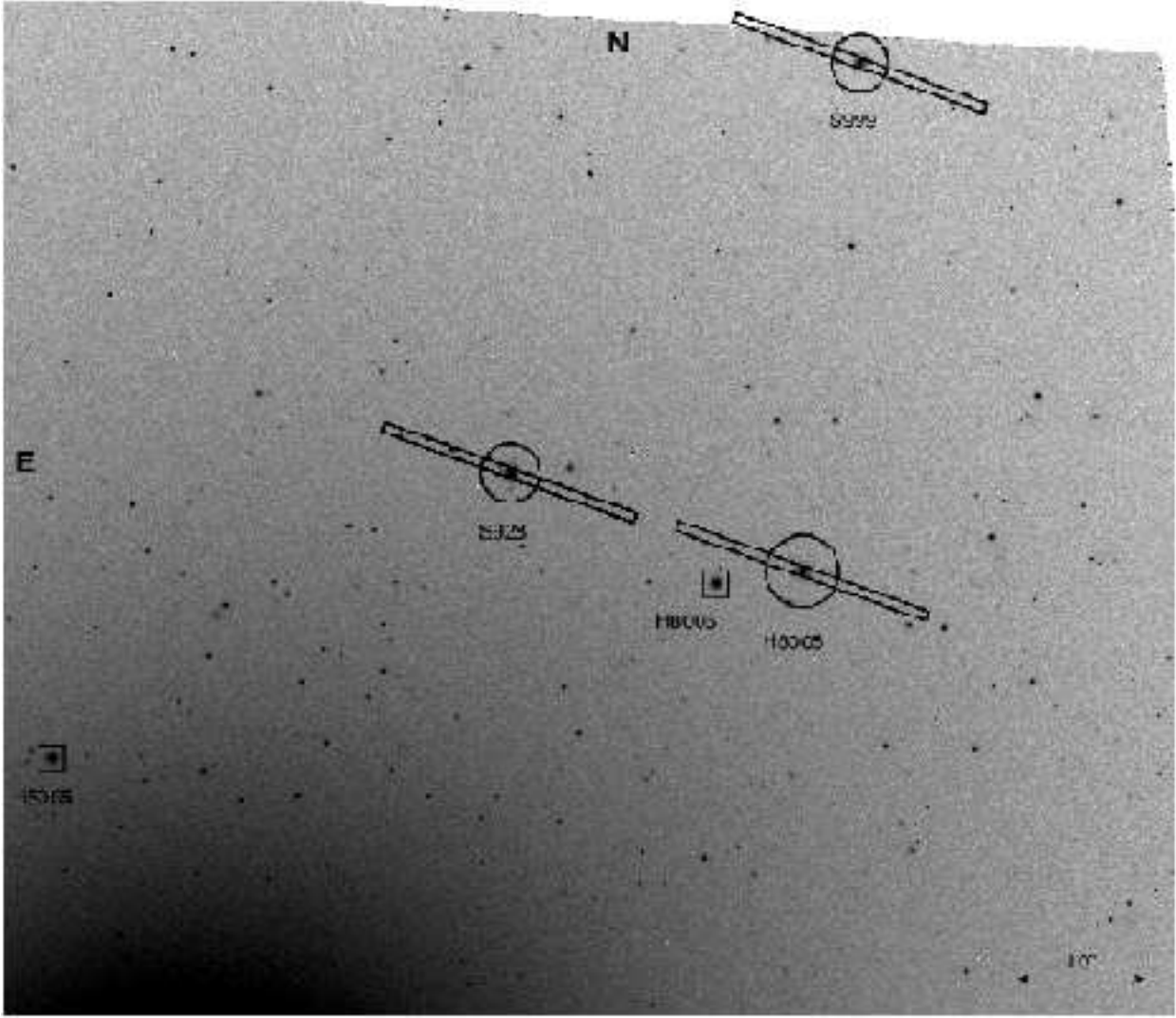}
\caption{F475W image of M87 from the ACS Virgo Cluster Survey, showing
the location of three DGTOs observed with Keck/ESI: S928, S999 and
H8005. The circles show the inferred tidal radii of each object.  The
0.75$\arcsec\times20\arcsec$ ESI slit used for the Keck observations
described in \S\ref{sec:spectr} is indicated by the rectangles. Two
additional DGTOs in this field, H5065 and H8006, are marked by the
squares (see \S8 for details).
\label{fig02}}
\end{figure}

Among the $\sim$ 7$\times10^4$ sources in the catalog, 27 objects
satisfied the criteria specified by Equations~\ref{sel1} and
\ref{sel2}. Each of these objects was inspected visually. Nine were
found to be obvious background galaxies or spurious detections such as
small-scale enhancements in rings or shells.  The 18 remaining objects
--- which include the three DGTOs from Figure~\ref{fig02} --- we
consider to be DGTO candidates. They are shown in the $r_h$-magnitude
planes in Figure~\ref{fig12} as the large squares where we have now
assumed a common distance of 16.5~Mpc for all galaxies. Small symbols
in this figure show all sources from the ACS Virgo Cluster Survey.
Generally speaking, three types of sources appear in this figure:
Galactic field stars with $r_h \lesssim 1$~pc; globular clusters with
mean $r_h \sim$ 2-3~pc, $21 \lesssim g \lesssim 26$ and $20 \lesssim z
\lesssim 25$; and background galaxies, which occupy a diagonal
sequence of faint objects beginning at $r_h \sim 2$~pc and extended up
to $r_h \gtrsim 200$~pc.  The dashed region indicates the selection
criteria given by Equation~\ref{sel1}.

The properties of the 18 DGTO candidates plotted in this figure are
summarized in Table 9, whose first column records an identification
number composed of the VCC number of the host galaxy and a counter
specifying an order number for candidates residing in the same galaxy.
Columns 2--9 give other names for these objects, their coordinates,
magnitudes, colors, half-light radii and concentration indices. The
final three columns record any published radial velocity measurements
(either from Table~4, \citealt{hanes01}, or \citealt{cote03}),
distance moduli from our own surface brightness fluctuation (SBF)
measurements (see below), and comments on their nature.

Figure~\ref{fig13} and \ref{fig14} show the appearance of the 18 DGTOs
in our F475W images.  The line in the first panel of each figure has a
length of 2\arcsec, highlighting their compact nature. Although our
selection criteria should eliminate the majority of background
galaxies, it is likely that at least some of these objects are
early-type galaxies at high and intermediate redshift. To gauge the
extent of this contamination, we have searched the object catalogs
generated from 17 control fields for the ACS Virgo Cluster Survey,
which are based on images collected from the {\it HST} archive and
reduced with the same reduction pipeline used for the survey itself
(see Paper II).  Observations in these fields were carried out using
the same filters as the Virgo program, and the combined images have
comparable depths and drizzling strategies (see Peng et~al. 2005 for
details). Selecting objects on the basis of Equations~\ref{sel1} and
\ref{sel2}, we find only two objects which would qualify as DGTO
candidates, one of which can immediately be classified as a background
galaxy based on its visual appearance. Since our control fields cover
an area (17/100) $\approx$ 1/6th that of the Virgo fields, we expect
$\sim$ 6 of the DGTO candidates listed in Table~9 to be background
galaxies.

It is possible to determine which of the 18 candidates in Table~9 are
 true DGTOs by using additional information, such as radial velocities
 and/or distances, to test for membership in Virgo. For seven objects
 in Table~9, radial velocities are available either from the
 literature or from this study.  Six of these objects, which were
 included in radial velocity surveys of the globular cluster systems
 of VCC1226 \citep[M49;][]{cote03} and VCC1316
 \citep[M87;][]{hanes01,cote01}, can be classified unambiguously as
 members of the Virgo Cluster. The redshift of the seventh object
 (VCC1316\_1 = S1063) reveals it to be a definite background galaxy
 \citep{huchra87}.

An alternative means of establishing membership is through distances.
The ACS Virgo Cluster Survey has been designed, in part, to provide
accurate SBF distances for our program galaxies (see
\citealt{mei05a,mei05b}; Papers IV and V). Needless to say, applying
the SBF technique to the DGTOs candidates is challenging because of
their compact sizes (which render the observable fluctuations to
spatial scales comparable to the PSF), the contamination of their
fluctuations from the adjacent galaxy, and the statistical
uncertainties involved in calibrating their stellar
populations. Nevertheless, we were able to measure crude SBF
magnitudes, $\overline m_z$, for 14 of the DGTO candidates in
Table~9. Although the errors on these measurements are of order
0.5~mag, they are sufficient to discriminate between the competing
scenarios of a dwarf galaxy or globular cluster residing in Virgo and
a giant elliptical lying well in the background.  For instance, $28
\lesssim \overline m_{z} \lesssim 29$ for dwarf ellipticals in Virgo,
whereas $\overline m_{z}$ for Virgo giants fall in the range
29--30~mag (Paper V). Thus, candidates with fluctuation magnitudes
$\overline m_{z} > 30$ can be classified with some confidence as
non-members of Virgo, although in such cases, the $\overline m_z$
measurements should be considered lower limits.  It is reassuring that
memberships established from radial velocities and SBF measurements
are in agreement in every case where both methods could be applied
(VCC1316\_1, VCC1316\_2, VCC1316\_3 and VCC1316\_4).

Considering both the radial velocity and SBF measurements, we conclude
that nine of the 18 objects in Table~9 are {\it certain} members of
Virgo; four additional objects are classified as {\it possible}
members. We consider the remaining five objects to be certain or
probable background galaxies. Figure~\ref{fig15} shows the color
distribution of all 18 objects. It is clear that the confirmed DGTOs
(i.e., those classified as certain members of Virgo) comprise a rather
homogeneous population in terms of color, with 0.88~$<(g-z)_0
<$~1.18. If these DGTOs are old (i.e., with ages greater than a few
Gyr), then their colors point to low to intermediate metallicities:
[Fe/H] $\lesssim -1$~dex (see Figure~5 of Paper I). This estimate is
consistent with the generally low metallicities found for S999, S928
and H8005 from independent photometry and spectroscopy (see Table~1).
In any case, Figure~\ref{fig15} suggests that the relatively narrow
color range of the certain DGTOs may be used to guide the membership
classifications for those objects lacking radial velocities and SBF
measurements.  The final column of Table~9 summarizes our conclusions
regarding the nature of each object, and the criteria used to
establish membership (i.e., radial velocities, SBF measurements, or
colors).

Three of the 13 certain or possible DGTOs merit special attention:
VCC1250\_1, VCC1297\_1 and VCC1632\_1. Figure~\ref{fig16}, reveals
that each of these objects has a surrounding halo of diffuse
light. These objects appear to be similar to UCD3, the brightest of
the compact objects in Fornax, which \citet{drink01} have suggested is
surrounded by a low-surface brightness envelope.  Unfortunately, for
none of the three Virgo objects is a radial velocity
available. Moreover, the ${\overline m_z}$ measurements for VCC1250\_1
and VCC1297\_1 are $\sim$ 1~mag larger than one might expect for old,
metal-poor dwarfs within Virgo. The significance of this result,
however, is unclear given the large measurement errors. In addition,
if they are really distant background galaxies, then it would be
surprising to find colors which so closely resemble those of the
confirmed DGTOs.  We conclude that a definitive conclusion regarding
membership is not possible for these two objects. On the other hand,
the SBF distance for VCC1632\_1 clearly points to membership in
Virgo. It is perhaps worth noting that, with $(g-z)_0 \approx 1.18$,
it would be the reddest of the confirmed DGTOs in our sample. Clearly,
spectroscopy for all three of these objects is urgently needed; if
they are truly Virgo members, their diffuse envelopes would provide
irrefutable evidence of a direct link between the nuclei of
low-surface brightness dwarfs and at least some DGTOs.  Moreover, the
diffuse light surrounding VCC1632\_1 is noteworthy in that the
intensity contours become increasingly flattened with radius, and may
even show a hint of isophotal twisting. If this is the case, then
VCC1632\_1 may be a DGTO that is currently forming through the tidal
stripping of a nucleated dwarf galaxy.  Spectroscopy of this object
would be extremely interesting: to settle the issue of membership and,
if it is found to reside in Virgo, to search for the ordered motions
that are expected to accompany tidal stripping \citep{piatek95}.

The nine certain DGTOs in Table~9 are associated with four different
galaxies: VCC798 (M85) (2), VCC1226 (M49) (1), VCC1316 (M87) (5) and
VCC1632 (M89) (1).  It is remarkable that five of the objects, or 56\%
of the sample, are associated with a single galaxy (M87). If we
consider both the certain and possible DGTOs, the overabundance still
remains: 5/13, or $\approx 38\%$ of the sample. But how surprising is
this excess?  After all, M87 is unique among Virgo galaxies in several
respects: it is one of the brightest members of Virgo, it has both a
remarkably rich globular cluster system \citep{mclaugh94} and a cD
envelope \citep{vaucoul78}, it is located close to Virgo's center as
traced by the cluster X-ray emission \citep{goren77}, and it is
roughly centered inside the ``core" of the galaxy surface density
profile \citep{bingg87}.  To understand how the properties of M87
(along with the three other galaxies containing confirmed DGTOs) may
promote or inhibit the formation of DGTOs, we compare in
Figure~\ref{fig17} their fraction of the total DGTO population,
$\eta_{\rm UCD}$, to those for other tracer components:

\begin{itemize}
\item[1.] $\eta_{\rm lum} \equiv {\cal L_B/L_{B,{\rm T}}}$: The
fraction of the blue luminosity contributed by each galaxy, ${\cal
L_{B}}$, to the total of the ACS Virgo Cluster Survey sample, ${\cal
L_{B,{\rm T}}} \approx 6.4\times10^{11}{\cal L_{B,\odot}}$.

\item[2.] $\eta_{\rm gc} \equiv N_{\rm gc}/N_{\rm gc,T}$: The fraction
of the total globular cluster population from the ACS Virgo Cluster
Survey, $N_{\rm gc,T} \approx$ 1.3$\times$10$^4$, that is associated
with an individual galaxy, $N_{\rm gc}$. We consider globular clusters
to be those objects with probability indices, ${\cal P_{\rm gc}}$, in
the range $0.5 \le {\cal P_{\rm gc}} \le 1$ (see Peng et~al. 2005 for
details).

\item[3.] $\eta_{\rm gal} \equiv N_{\rm gal}/N_{\rm gal,T}$: The
number of early-type member galaxies, $N_{\rm gal}$, from the Virgo
Cluster Catalog \citep{bingg85} which have $B_T \ge 14$ and are
located within 1\fdg5~of each galaxy, normalized by the total number
of such galaxies in Virgo, $N_{\rm gal,T} = 889$.\footnote{Since
VCC798 lies $\sim$ 1/3 deg. from the boundary of region surveyed by
\citet{bingg85}, the calculated value of $\eta_{\rm gal}$ for this
galaxy should be viewed as a lower limit.}
\end{itemize}

Figure~\ref{fig17} reveals that the number of DGTOs associated with M87
far exceeds that expected based on its luminosity, the size of its
globular cluster system, and the local surface density of dwarf
galaxies. For instance, M49 is comparably luminous ($\eta_{\rm lum} =
0.13$ versus 0.10 for M87) and its globular cluster system is about
half as large as that of M87.  Nevertheless, the number of DGTOs
associated with M87 is five times that for M49. We suspect that unique
location of M87 --- i.e., sitting at the bottom of the gravitational
potential well defined by Virgo and itself, and near the centroid of 
the dwarf galaxy population in Virgo; \cite{bingg87,cote01} --- must
play a role in the origin of this excess.  However, it seems clear
that the total number of DGTOs is modest compared to the enormous M87
globular cluster system, and that they have virtually no effect on its
overall properties such as specific frequency.

Finally, we note a curious result from our search for DGTOs in
Virgo. Five of the nine certain DGTOs --- selected from thousands of
objects in 100 fields covering an area of $\approx$ 1100 arcmin$^{2}$
--- are located in a single $\approx$ 2 arcmin$^{2}$ region within the
northwest quadrant of the M87 field. This field happens to lie along
the ``principal axis" of the Virgo Cluster, which follows the major
axis of M87 (and its globular cluster system) on small scales and is
defined by the large-scale distribution of galaxies further out
\citep{bingg87,west00}.  Kinematical studies of Virgo/M87, focussing
on its dwarf galaxies and X-ray-emitting gas \citep{bingg99} and
globular clusters \citep{cote01}, suggest that the population of dwarf
galaxies in Virgo is probably not yet in equilibrium, with galaxies
currently infalling along the principal axis. The DGTOs in M87 may
thus provide indirect evidence for an evolutionary link to the
predominantly low- and intermediate-luminosity galaxies which define
this structure.

\section{Summary}
\label{sec:conclusions}

Motivated by the discovery of a population of faint, compact objects
in the Fornax Cluster, we have carried out a detailed study of similar
objects in Virgo, selected from ground-based studies of the M87
globular cluster system and imaging from the ACS Virgo Cluster Survey.
In terms of luminosity, the ``dwarf globular transition objects" which
are the focus of this study occupy a luminosity regime which is
populated by the brightest known globular clusters and the faintest dE
galaxies. While it is perhaps not surprising that we find our sample
to include {\it both} globular clusters and UCDs, our analysis has
shown that extreme care must be exercised to distinguish UCDs from
globular clusters in this luminosity range.

Among the six DGTOs in our sample which have measured dispersions,
structural parameters and mass-to-light ratios, we find two to be
massive, but otherwise normal, globular clusters.  The four remaining
objects we classify as probable or possible UCDs based on their
metallicities, large sizes, and higher mass-to-light ratios. This
latter parameter is found to be a particularly powerful tool for
identifying UCDs. In terms of mass-to-light ratio, our UCDs resemble
the nuclei of dE,N galaxies in Virgo, for which \citet{geha02} have
measured ${\cal M/L}_V \approx$ 5. Assuming the fundamental assumption
of virial equilibrium to be valid for the UCDs and dwarf nuclei, their
mass-to-light ratios cannot be explained by simple stellar population
differences. Rather, it appears that dark matter halos are needed to
account for the measured mass-to-light ratios.

Extrapolating down from the scales of luminous elliptical galaxies, we
find both the dwarf nuclei and UCDs to obey the galactic scaling
relations (i.e., those involving mass, central velocity dispersion,
half-light radius and mass surface density).  This suggests a
connection between the two populations, most obviously through the
removal of low-surface brightness envelopes in dE,Ns by tidal
stripping. Moreover, if the principal characteristic that
distinguishes UCDs and dwarf nuclei from globular clusters is the
presence of a dark matter halo, then this would naturally lead to a
different set of scaling relations.  A search for additional DGTOs in
the ACS Virgo Cluster Survey has revealed 13 certain or possible DGTO
candidates; three of these objects show some evidence for diffuse
envelopes, although in no case can membership be established
unambiguously at this time.  Taken together, the assembled evidence is
consistent with the formation of at least some UCDs through galaxy
threshing \citep{bekki01}. At the same time, one UCD in our sample
(H8005) is found to have properties which agree remarkably well with
the predictions of UCD formation through multiple mergers of young
massive star clusters \citep{fellhauer02}. This suggests that there
may well exist multiple formation routes for UCDs.

Certainly the overabundance of DGTOs associated with M87 --- which
contains about half of the total number of DGTOs uncovered in our
survey of Virgo galaxies --- suggests that proximity to the
gravitational center of Virgo must play a key role in their
formation. Remarkably, five of the 13 DGTOs uncovered in our survey of
Virgo happen to lie within a single $\approx$ 2 arcmin$^2$ region in
the northwest quandrant of our M87 field. Keck spectroscopy is
available for three of these five, and an analysis of their structural
and dynamical properties reveals each of them to be a probable or
possible UCD. It seems clear, however, that UCDs make up only a minute
portion of the M87 globular cluster system and have virtually no
effect on its overall properties.
  
Some obvious followup studies present themselves. Spectroscopy for our
sample of candidate DGTOs in Virgo would be desirable, first and
foremost to establish membership, but also to obtain velocity
dispersions, metallicities and mass-to-light ratios needed to
distinguish UCDs from globular clusters. In addition, the detection of
low-surface brightness material surrounding UCDs would help bolster
the case for an evolutionary link with nucleated galaxies.  In this
regard, new and better photometric, structural and dynamical data for
dE,N galaxies and their nuclei would provide an invaluable point of
comparison.

\acknowledgments

Support for program GO-9401 was provided through a grant from the
Space Telescope Science Institute, which is operated by the
Association of Universities for Research in Astronomy, Inc., under
NASA contract NAS5-26555.  P.C. acknowledges additional support
provided by NASA LTSA grant NAG5-11714.  A.J. acknowledges additional
financial support provided by the National Science Foundation through
a grant from the Association of Universities for Research in
Astronomy, Inc., under NSF cooperative agreement AST-9613615, and by
Fundaci\'on Andes under project No.C-13442.  M.M. acknowledges
additional financial support provided by the Sherman M. Fairchild
foundation. D.M. is supported by NSF grant AST-020631, NASA grant
NAG5-9046, and grant HST-AR-09519.01-A from STScI.
M.J.W. acknowledges support through NSF grant AST-0205960.
S.G.D. acknowledges a partial support from the Ajax Foundation, and
the creative atmosphere of the Aspen Center for Physics where some of
this work was done. We thank Milan Bogosavljevic for help with ESI
data, and the staff of W.M. Keck Observatory for their expert
assistance during our observing runs.  We are grateful to P.B. Stetson
for the F555W PSF and to Michael Drinkwater for the Fornax UCD data.
This research has made use of the NASA/IPAC Extragalactic Database
(NED) which is operated by the Jet Propulsion Laboratory, California
Institute of Technology, under contract with the National Aeronautics
and Space Administration.  This work was based on observations
obtained at the W.M. Keck Observatory, which is operated jointly by
the California Institute of Technology and the University of
California. We are grateful to the W.M. Keck Foundation for their
vision and generosity. Finally, we recognize the great importance of
Mauna Kea to both the native Hawaiian and astronomical communities,
and we are grateful for the opportunity to observe from this special
place.

\clearpage

\begin{deluxetable}{lccccccccccc}
\tablecaption{Coordinates, Photometry and Metallicities for DGTOs\label{tab1}}
\scriptsize
\rotate
\tabletypesize{\scriptsize}
\tablewidth{0pt}
\tablehead{
\colhead{ID} &
\colhead{$\alpha$(J2000)} &
\colhead{$\delta$(J2000)} &
\colhead{$R$} &
\colhead{$V$} &
\colhead{$g$} &
\colhead{$z$} &
\colhead{$(g-z)_0$} &
\colhead{($C-T_1$)} &
\colhead{[Fe/H]$_{gz}$} &
\colhead{[Fe/H]$_{CT_1}$}&
\colhead{[Fe/H]$_{\rm CBR}$} \\
\colhead{} &
\colhead{} &
\colhead{} &
\colhead{(arcsec)} &
\colhead{(mag)} &
\colhead{(mag)} &
\colhead{(mag)} &
\colhead{(mag)} &
\colhead{(mag)} &
\colhead{(dex)} &
\colhead{(dex)} &
\colhead{(dex)}
}
\startdata
 S314 & 12:31:05.10 & +12:20:04.0 & 307.2 & 20.19 & \nodata & \nodata  & \nodata   & 1.68 & \nodata & $-$0.50 &$-$0.34 \\
 S348 & 12:31:03.40 & +12:23:06.0 & 206.0 & 19.61 & \nodata & \nodata  & \nodata   & 1.30 & \nodata & $-$1.36 &$-$1.25 \\
 S417 & 12:31:01.50 & +12:19:25.0 & 300.6 & 19.33 & \nodata & \nodata  & \nodata   & 1.59 & \nodata & $-$0.70 &$-$0.61 \\
 S490 & 12:30:59.30 & +12:21:23.0 & 191.3 & 20.11 & \nodata & \nodata  & \nodata   & 1.98 & \nodata & $+$0.18 &$+$0.11 \\
 S804 & 12:30:51.10 & +12:26:12.0 & 165.8 & 19.29 & \nodata & \nodata  & \nodata   & 1.55 & \nodata & $-$0.79 & \nodata\\
 S928 & 12:30:47.70 & +12:24:30.8 &  67.7 & 19.57 & 19.960  & 18.977   & 0.936     & 1.31 & $-$1.54 & $-$1.34 &$-$1.37 \\
 S999 & 12:30:45.91 & +12:25:01.8 & 107.0 & 20.03 & 20.332  & 19.402   & 0.883     & 1.29 & $-$1.93 & $-$1.38 & \nodata\\
S1370 & 12:30:37.40 & +12:19:18.0 & 305.8 & 19.63 & \nodata & \nodata  & \nodata   & 1.60 & \nodata & $-$0.68 &$-$0.49 \\
S1538 & 12:30:30.60 & +12:22:56.0 & 277.6 & 19.83 & \nodata & \nodata  & \nodata   & 1.66 & \nodata & $-$0.54 &$-$0.61 \\
H8005 & 12:30:46.21 & +12:24:23.3 &  72.6 & 20.27 & 20.602  & 19.603   & 0.952     & 1.34 & $-$1.45 & $-$1.27 & \nodata\\
\enddata
\tablenotetext{~}{Notes:
-- Units of right ascension are hours, minutes, and seconds, and units of declination 
are degrees, arcminutes, and arcseconds.\\
}

\end{deluxetable}

\clearpage

\begin{deluxetable}{lccccc}
\tablecaption{ESI Observing Log for DGTOs\label{tab2}}
\scriptsize
\tablewidth{0pt}
\tablehead{
\colhead{ID} &
\colhead{$T$} &
\colhead{HJD} &
\colhead{$\Theta$} &
\colhead{FWHM} &
\colhead{$H_x$} \\
\colhead{} &
\colhead{(sec)} &
\colhead{} &
\colhead{(deg)} &
\colhead{(arcsec)} &
\colhead{(arcsec)}
}
\startdata
 S314 & 1800 & 2451664.77054  & 290 & 0.70 & 0.77 \\
 S348 & 1500 & 2451664.79356  & 295 & 0.73 & 0.82 \\
 S417 & 1500 & 2451665.73963  & 290 & 1.14 & 1.18 \\
 S490 & 1800 & 2451664.81370  & 305 & 0.86 & 0.89 \\
 S804 & 1500 & 2451664.75035  & 290 & 0.87 & 1.04 \\
 S928 & 1800 & 2452794.82831  &  70 & 0.97 & 1.18 \\
 S999 & 1200 & 2452794.81090  &  70 & 1.18 & 1.83 \\
      & 1200 & 2452794.85367  &  70 & 1.10 & 1.39 \\
S1370 & 1800 & 2451665.75923  & 285 & 0.99 & 0.96 \\
S1538 & 1800 & 2451665.78279  & 295 & 1.08 & 1.20 \\
H8005 & 1800 & 2452794.87004  &  70 & 1.13 & 1.14 \\
      & 1100 & 2452794.89157  &  70 & 1.16 & 1.15 \\
\enddata

\end{deluxetable}

\clearpage
\begin{deluxetable}{lccccccc}
\tablecaption{ESI Observing Log for Radial Velocity Standard Stars\label{tab3}}
\scriptsize
\tablewidth{0pt}
\tablehead{
\colhead{ID} &
\colhead{$\alpha$(2000)} &
\colhead{$\delta$(2000)} &
\colhead{$T$} &
\colhead{HJD} &
\colhead{Type} &
\colhead{$V$} &
\colhead{$v_r$}  \\
\colhead{} &
\colhead{} &
\colhead{} &
\colhead{(sec)} &
\colhead{} &
\colhead{} &
\colhead{(mag)} &
\colhead{(km s$^{-1}$)}
}
\startdata
HD86801  & 10:01:36.13 & +28:34:02.3 & 20    & 2451664.73377 & G0 V      & 8.78 & $-$14.5 \\
HD102494 & 11:47:57.13 & +27:20:16.2 &  6    & 2452794.80510 & G9 IV      & 7.48 & $-$22.9\\
HD107328 & 12:20:22.59 & +03:18:37.7 &  1    & 2452794.80438 & K0.5 III  & 4.96 & +35.7\\
HD112299 & 12:55:29.08 & +25:44:05.9 & 15    & 2451664.74060 & F8 V      & 8.39 &  +3.4 \\
HD132737 & 14:59:54.04 & +27:09:30.2 &  8    & 2452794.91145 & K0 III     & 7.64 & $-$24.1\\
HD145001 & 16:08:06.18 & +17:02:35.3 &  1    & 2451665.01091 & G5 III    & 5.00 & $-$9.5\\
HD154417 & 17:05:18.85 & +00:42:11.1 &  1    & 2451665.01841 & F8.5 IV-V & 6.01 & $-$17.4\\
HD194071 & 20:22:38.94 & +28:14:54.6 &  5    & 2451665.13127 & G8 III    & 7.80 &  $-$9.8\\
HD203638 & 21:24:11.25 & $-$20:50:53.4 &  1  & 2451665.14021 & K0 III    & 5.41 & +21.9\\

\enddata
\tablenotetext{~}{Note.-- Units of right ascension are hours, minutes, and seconds, and units of declination 
are degrees, arcminutes, and arcseconds.}
\end{deluxetable}

\clearpage

\begin{deluxetable}{lccccccccccccccc}
\tablecaption{Radial Velocities and Velocity Dispersions for DGTOs with ACS Imaging\label{tab4}}
\rotate
\tabletypesize{\scriptsize}
\tablewidth{0pt}
\tablehead{
 & & \multicolumn{5}{c}{$\sigma$(FXCOR)} & & \multicolumn{5}{c}{$\sigma$(pPXF)} & \\
\cline{3-7} \cline{9-13} 
\colhead{ID} &
\colhead{$v_{r}$} &
\colhead{A} &
\colhead{B} &
\colhead{C} &
\colhead{D} &
\colhead{E} &
\colhead{$\langle\sigma\rangle$} &
\colhead{A} &
\colhead{B} &
\colhead{C} &
\colhead{D} &
\colhead{E} &
\colhead{$\langle\sigma\rangle$} &
\colhead{$\sigma_{0}$} \\
\colhead{} &
\colhead{(km~s$^{-1}$)} &
\colhead{} &
\colhead{} &
\colhead{(km~s$^{-1}$)} &
\colhead{} &
\colhead{} &
\colhead{(km~s$^{-1}$)} &
\colhead{} &
\colhead{} &
\colhead{(km~s$^{-1}$)} &
\colhead{} &
\colhead{} &
\colhead{(km~s$^{-1}$)} &
\colhead{(km~s$^{-1}$)} 
}
\startdata
S928 & 1282.5$\pm$5.0 & 22.6 & 22.2 & 21.7 & 21.4 & 21.4 & 21.9$\pm$0.5 & 19.8$\pm$0.4 & 19.4$\pm$0.5 & 19.1$\pm$0.5 & 21.3$\pm$0.3 & 20.4$\pm$0.4 & 20.0$\pm$0.9 & 22.4$\pm$1.0\\
S999 & 1465.9$\pm$5.1 & 18.6 & 19.6 & 18.9 & 21.3 & 19.6 & 19.6$\pm$1.0 & 22.4$\pm$0.3 & 22.5$\pm$0.4 & 22.3$\pm$0.3 & 25.2$\pm$0.3 & 24.3$\pm$0.4 & 23.3$\pm$1.3 & 25.6$\pm$1.4\\
H8005 & 1882.6$\pm$5.1& 11.4 &  7.0 &  9.5 & 12.3 &  9.7 & 9.98$\pm$2.0 & 10.1$\pm$1.5 &  7.6$\pm$1.8 &  7.1$\pm$0.5 & 11.8$\pm$0.3 &  8.7$\pm$0.4 & 9.1$\pm$1.9 & 10.8$\pm$2.3\\
\enddata
\tablenotetext{~}{Key to templates: 
(A) HD107328;
(B) HD102494;
(C) HD102494;
(D) HD132737;
(E) HD132737.}

\end{deluxetable}

\begin{deluxetable}{lccccccccccccccccc}
\tablecaption{Radial Velocities and Velocity Dispersions for DGTOs without ACS Imaging\label{tab5}}
\rotate
\tabletypesize{\scriptsize}
\tablewidth{0pt}
\tablehead{
 & & \multicolumn{6}{c}{$\sigma$(FXCOR)} & & \multicolumn{6}{c}{$\sigma$(pPXF)} & \\
\cline{3-8} \cline{10-15} 
\colhead{ID} &
\colhead{$v_{r}$} &
\colhead{A} &
\colhead{B} &
\colhead{C} &
\colhead{D} &
\colhead{E} &
\colhead{F} &
\colhead{$\langle\sigma\rangle$}& 
\colhead{A} &
\colhead{B} &
\colhead{C} &
\colhead{D} &
\colhead{E} &
\colhead{F} &
\colhead{$\langle\sigma\rangle$} &
\colhead{$\sigma_{0}$} \\
\colhead{} &
\colhead{(km~s$^{-1}$)} &
\multicolumn{6}{c}{(km~s$^{-1}$)} &
\colhead{(km~s$^{-1}$)} &
\multicolumn{6}{c}{(km~s$^{-1}$)} &
\colhead{(km~s$^{-1}$)} &
\colhead{(km~s$^{-1}$)}
}
\startdata
 S314\tablenotemark{a} & 1220.6$\pm$10.1 & 37.3 & 36.4 & 36.9 & 39.4 & 37.1 & 36.2 & 37.2$\pm$1.1 & 35.5$\pm$0.2 & 35.3$\pm$0.2 & 36.2$\pm$0.2 & 35.5$\pm$0.2 & 36.2$\pm$0.2 & 32.4$\pm$0.2 & 35.2$\pm$1.4 & 35.3 $\pm$ 1.4 \\
 S348 &  795.7$\pm$10.1 & 25.1 & 23.8 & 26.8 & 27.8 & 25.7 & 26.2 & 25.9$\pm$1.4 &20.8$\pm$0.3 &19.8$\pm$0.3 & 23.1$\pm$0.2 & 21.5$\pm$0.2 & 22.6$\pm$0.3 & 20.4$\pm$0.2 & 21.4$\pm$1.3 & \nodata \\
 S417\tablenotemark{a} & 1862.1$\pm$4.2 & 31.5 & 31.5 & 32.7 & 31.7 & 32.6 & 32.6 & 32.1$\pm$0.6 & 30.9$\pm$0.2 & 30.0$\pm$0.2 & 31.9$\pm$0.2 & 31.0$\pm$0.2 & 31.6$\pm$0.2 & 27.7$\pm$0.2 & 30.5$\pm$1.5 & 32.9 $\pm$ 1.6 \\
 S490\tablenotemark{a} & 1566.8$\pm$7.0 & 37.7 & 34.7 & 33.9 & 36.7 & 36.4 & 37.2 & 36.1$\pm$1.5 & 44.1$\pm$0.2 & 44.9$\pm$0.2 & 41.5$\pm$0.2 & 42.3$\pm$0.2 & 42.6$\pm$0.2 & 37.1$\pm$0.2 & 42.1$\pm$2.7 & 42.5 $\pm$ 2.7 \\
 S804 &	1134.3$\pm$10.0 & 34.5 & 33.6 & 35.8 & 36.6 & 34.8 & 36.5 & 35.3$\pm$1.2 & 33.4$\pm$0.2 & 32.8$\pm$0.2 & 34.9$\pm$0.2 & 33.3$\pm$0.2 & 34.6$\pm$0.2 & 32.1$\pm$0.2 & 33.5$\pm$1.1 & \nodata \\
 S1370 & 1083.9$\pm$4.1 & 39.8 & 39.1 & 40.6 & 40.3 & 40.0 & 39.2 & 39.8$\pm$0.6 & 38.6$\pm$0.2 & 38.8$\pm$0.2 & 39.1$\pm$0.2 & 38.6$\pm$0.2 & 39.2$\pm$0.2  & 34.8$\pm$0.2 & 38.2$\pm$1.7 & \nodata \\
 S1538 & 1196.9$\pm$4.2 & 25.8 & 28.8 & 30.4 & 31.6 & 33.6 & 29.5 & 30.0$\pm$2.6 & 26.3$\pm$0.2 & 26.2$\pm$0.2 & 27.6$\pm$0.2 & 26.8$\pm$0.2 & 27.5$\pm$0.2 & 24.9$\pm$0.2 & 26.6$\pm$1.0 & \nodata \\
\enddata
\tablenotetext{~}{Key to templates: 
(A) HD86801;
(B) HD112299;
(C) HD145001;
(D) HD154417;
(E) HD194071;
(F) HD203638.}
\tablenotetext{a}{Archival WFPC2 imaging available. See \S\ref{sec:wfpc2} and Table~\ref{tab7}.}

\end{deluxetable}

\clearpage

\begin{deluxetable}{lccccccccccccccc}
\tablecaption{Structural Parameters for DGTOs with ACS Imaging\label{tab6}}
\rotate
\tabletypesize{\scriptsize}
\tablewidth{0pt}
\setlength{\tabcolsep}{0.03in}
\tablehead{
\colhead{ID} &
\colhead{$\epsilon$} &
\colhead{$c_g$} &
\colhead{$c_z$} &
\colhead{$r_{h,g}$} &
\colhead{$r_{h,z}$} &
\colhead{$r_{c,g}$} &
\colhead{$r_{c,z}$} &
\colhead{$\mu_{V,g}$} &
\colhead{$\mu_{V,z}$} &
\colhead{$p_{g}$} &
\colhead{$p_{z}$} &
\colhead{$\nu_{g}$} &
\colhead{$\nu_{z}$} &
\colhead{$\alpha_{g}$} &
\colhead{$\alpha_{z}$} \\
\colhead{} &
\colhead{} &
\colhead{} &
\colhead{} &
\colhead{(pc)} &
\colhead{(pc)} &
\colhead{(pc)} &
\colhead{(pc)} &
\colhead{(mag/} &
\colhead{(mag/} &
\colhead{} &
\colhead{} &
\colhead{} &
\colhead{} &
\colhead{} & 
\colhead{} \\
\colhead{} &
\colhead{} &
\colhead{} &
\colhead{} &
\colhead{}&
\colhead{}&
\colhead{}&
\colhead{}&
\colhead{arcsec$^2$)} &
\colhead{arcsec$^2$)} &
\colhead{}&
\colhead{}&
\colhead{}&
\colhead{}&
\colhead{}&
\colhead{} 
}
\startdata
 S928 & 0.133 & 1.080$\pm$0.029 & 1.177$\pm$0.049 & 21.79$\pm$0.46 & 24.52$\pm$0.74 & 13.52$\pm$0.74 & 13.59$\pm$1.19 & 18.72 & 18.97 & 1.82$\pm$0.01 & 1.85$\pm$0.02 & 12.89$\pm$0.63 & 15.04$\pm$1.13 & 0.895$\pm$0.006 & 0.912$\pm$0.008\\
 S999 & 0.057 & 1.010$\pm$0.045 & 1.099$\pm$0.037 & 19.15$\pm$0.19 & 21.10$\pm$0.49 & 12.90$\pm$1.03 & 12.81$\pm$0.87 & 18.89 & 19.11 & 1.79$\pm$0.02 & 1.82$\pm$0.03 & 11.42$\pm$0.93 & 13.30$\pm$0.81 & 0.879$\pm$0.011 & 0.899$\pm$0.007 \\
H8005 & 0.171 & 1.303$\pm$0.069 & 1.304$\pm$0.071 & 28.14$\pm$1.64 & 29.24$\pm$2.37 & 13.39$\pm$1.70 & 13.90$\pm$1.97 & 19.98 & 20.06 & 1.88$\pm$0.02 & 1.88$\pm$0.02 & 18.12$\pm$1.80 & 18.14$\pm$1.86 & 0.930$\pm$0.009 & 0.930$\pm$0.009 \\
\enddata

\end{deluxetable}

\clearpage
\begin{deluxetable}{lccccccc}
\tablecaption{Adopted Structural Parameters for DGTOs\label{tab7}}
\scriptsize
\tablewidth{0pt}
\setlength{\tabcolsep}{0.06in}
\tablehead{
\colhead{ID} &
\colhead{$\langle c \rangle$} &
\colhead{$\langle r_{h} \rangle$} &
\colhead{$\langle r_{c} \rangle$} &
\colhead{$\langle \mu_{V}^h \rangle$} &
\colhead{$\langle p \rangle$} &
\colhead{$\langle \nu \rangle$} &
\colhead{$\langle \alpha \rangle$}\\
\colhead{} &
\colhead{} &
\colhead{(pc)} &
\colhead{(pc)} &
\colhead{(mag arcsec$^{-2}$)} &
\colhead{} &
\colhead{} & 
\colhead{} 
}
\startdata
S314 & 1.70$\pm$0.09 &  3.23$\pm$0.19 & 0.82$\pm$0.13  & 15.20 & 1.94$\pm$0.01 & 31.97$\pm$4.48 & 0.961$\pm$0.005 \\
S417 & 1.19$\pm$0.09 & 14.36$\pm$0.36 & 7.85$\pm$1.17  & 17.58 & 1.85$\pm$0.03 & 15.32$\pm$2.06 & 0.914$\pm$0.014 \\
S490 & 1.84$\pm$0.11 &  3.64$\pm$0.36 & 0.68$\pm$0.14  & 15.37 & 1.96$\pm$0.01 & 40.30$\pm$7.72 & 0.969$\pm$0.006 \\ 
S928 & 1.13$\pm$0.05 & 23.16$\pm$1.37 & 13.55$\pm$0.04 & 18.85 & 1.83$\pm$0.02 & 13.97$\pm$1.08 & 0.903$\pm$0.009 \\
S999 & 1.05$\pm$0.04 & 20.13$\pm$0.98 & 12.86$\pm$0.05 & 19.00 & 1.81$\pm$0.02 & 12.36$\pm$0.94 & 0.889$\pm$0.010 \\
H8005& 1.30$\pm$0.00 & 28.69$\pm$0.55 & 13.65$\pm$0.26 & 20.02 & 1.88$\pm$0.00 & 18.13$\pm$0.01 & 0.930$\pm$0.000 \\
\enddata

\end{deluxetable}

\clearpage

\begin{deluxetable}{lcccc}
\tablecaption{Masses and Mass-to-Light Ratios for DGTOs\label{tab8}}
\tablewidth{0pt}
\tablehead{
\colhead{ID} &
\colhead{$M_V$} &
\colhead{${\cal L}_V$} &
\colhead{${\cal M}_{k}$} &
\colhead{${\cal M}_{k}/{\cal L}_V$} \\
\colhead{} &
\colhead{(mag)} &
\colhead{($10^6{\cal L}_{V,{\odot}}$)} &
\colhead{($10^7{\cal M}_{{\odot}}$)} &
\colhead{(${\cal M}_{{\odot}}$/${\cal L}_{V,{\odot}}$)} 
}
\startdata
 S314 & $-$10.91$\pm$0.16 & 1.98$\pm$0.30 & 0.58$\pm$0.10 & 2.94$\pm$0.68 \\
 S417 & $-$11.78$\pm$0.16 & 4.39$\pm$0.66 & 2.56$\pm$0.46 & 5.83$\pm$1.36 \\
 S490 & $-$11.00$\pm$0.16 & 2.14$\pm$0.32 & 0.87$\pm$0.21 & 4.06$\pm$0.15 \\
 S928 & $-$11.58$\pm$0.16 & 3.52$\pm$0.53 & 2.13$\pm$0.29 & 6.06$\pm$1.23 \\
 S999 & $-$11.08$\pm$0.16 & 2.31$\pm$0.34 & 2.16$\pm$0.29 & 9.36$\pm$1.87 \\
H8005 & $-$10.83$\pm$0.16 & 1.84$\pm$0.28 & 0.55$\pm$0.23 & 2.98$\pm$1.35 \\
\enddata

\end{deluxetable}

\clearpage

\begin{deluxetable}{llcccccccccl}
\tablecaption{Data for DGTO Candidates in the ACS Virgo Cluster Survey\label{tab9}}
\rotate
\tabletypesize{\scriptsize}
\tablewidth{0pt}
\setlength{\tabcolsep}{0.04in}
\tablehead{
\colhead{ID} &
\colhead{Other\tablenotemark{a}} &
\colhead{$\alpha$(J2000)} &
\colhead{$\delta$(J2000)} &
\colhead{$g$} &
\colhead{$z$} &
\colhead{($g-z$)$_0$} &
\colhead{$\langle r_h\rangle$} &
\colhead{$\langle c \rangle$} &
\colhead{$v_r$\tablenotemark{b}} &
\colhead{$\overline m_{z}$} &
\colhead{Comments} \\
\colhead{} &
\colhead{} &
\colhead{} &
\colhead{} &
\colhead{(mag)} &
\colhead{(mag)} &
\colhead{(mag)} &
\colhead{(pc)} &
\colhead{} &
\colhead{(km s$^{-1})$} &
\colhead{(mag)} &
\colhead{}
}
\startdata
\multicolumn{12}{c}{\centerline{\it \underline{Certain or Probable DGTOs}}} \\
798\_1  &       & 12:25:20.980 & +18:10:12.57 & 20.978 & 19.962 & 0.951 & 14.52$\pm$0.70 & 1.21$\pm$0.06 &      &    28.2 & Certain (SBF, color)\\
798\_2  &       & 12:25:26.018 & +18:11:47.95 & 19.923 & 18.907 & 0.951 & 19.02$\pm$0.76 & 1.13$\pm$0.05 &      &    28.6 & Certain (SBF, color)\\
1226\_1 & G9992 & 12:29:48.347 & +08:00:41.79 & 20.886 & 19.718 & 1.120 & 18.86$\pm$1.54 & 1.07$\pm$0.12 &  795 & \nodata & Certain (velocity, color)\\
1316\_2 & S999  & 12:30:45.912 & +12:25:01.80 & 20.332 & 19.402 & 0.880 & 22.34$\pm$1.74 & 1.12$\pm$0.11 & 1515 &    28.7 & Certain (velocity, SBF, color), UCD\\
1316\_3 & H8005 & 12:30:46.209 & +12:24:23.30 & 20.602 & 19.603 & 0.949 & 26.47$\pm$0.93 & 1.17$\pm$0.06 & 1934 &    28.4 & Certain (velocity, SBF, color), UCD?\\
1316\_4 & H8006 & 12:30:46.653 & +12:24:22.40 & 20.570 & 19.565 & 0.955 & 21.61$\pm$1.37 & 1.18$\pm$0.09 & 1071 &    27.9 & Certain (velocity, SBF, color)\\
1316\_5 & S928  & 12:30:47.704 & +12:24:30.80 & 19.960 & 18.977 & 0.933 & 23.45$\pm$0.43 & 1.10$\pm$0.03 & 1299 & \nodata & Certain (velocity, color), UCD\\
1316\_6 & H5065 & 12:30:50.049 & +12:24:09.15 & 20.331 & 19.355 & 0.926 & 13.90$\pm$0.21 & 0.88$\pm$0.03 & 1563 & \nodata & Certain (velocity, color)\\
1632\_1 &       & 12:35:38.106 & +12:33:01.17 & 19.712 & 18.449 & 1.175 & 39.40$\pm$1.77 & 2.00$\pm$0.03 &      &    28.6 & Probable (SBF, color)\\
\multicolumn{12}{c}{\centerline{\it \underline{Possible DGTOs}}} \\
1250\_1 &       & 12:29:56.685 & +12:19:31.46 & 20.462 & 19.472 & 0.931 & 11.14$\pm$0.34 & 1.12$\pm$0.06 &      &    29.8 & Possible (SBF)\\
1297\_1 &       & 12:30:33.419 & +12:29:54.16 & 19.264 & 18.201 & 1.017 & 36.50$\pm$1.71 & 1.54$\pm$0.04 &      &    29.8 & Possible (SBF)\\
1632\_2 &       & 12:35:35.194 & +12:33:41.72 & 20.262 & 19.273 & 0.901 & 10.13$\pm$0.09 & 0.84$\pm$0.03 &      &    29.6 & Possible (SBF, color)\\
1861\_1 &       & 12:41:01.441 & +11:09:04.25 & 20.810 & 19.751 & 0.996 & 21.52$\pm$2.73 & 1.79$\pm$0.11 &      & \nodata & Possible (color)\\
\multicolumn{12}{c}{\centerline{\it \underline{Certain or Probable Background Galaxies}}} \\
763\_1  &       & 12:25:07.503 & +12:52:55.62 & 20.105 & 18.487 & 1.530 & 72.77$\pm$7.28 & 2.15\tablenotemark{c} &      &$\gtrsim$30.6 & Probable (SBF, color) \\
828\_1  &       & 12:25:44.093 & +12:49:22.14 & 20.244 & 19.135 & 1.038 & 22.03$\pm$1.10 & 1.11$\pm$0.08         &      &$\gtrsim$30.4 & Probable (SBF)\\
1316\_1 & S1063 & 12:30:45.630 & +12:22:12.44 & 20.699 & 19.089 & 1.560 & 41.91$\pm$1.88 & 2.04$\pm$0.03         &25396 &$\gtrsim$30.6 & Certain (velocity, SBF, color)\\
1627\_1 &       & 12:35:40.852 & +12:23:10.71 & 20.634 & 19.129 & 1.422 & 17.94$\pm$2.98 & 1.73$\pm$0.17         &      &$\gtrsim$28.8 & Probable (color, spiral morphology?)\\
1895\_1 &       & 12:41:51.700 & +09:22:52:55 & 19.528 & 17.906 & 1.585 & 92.27$\pm$8.17 & 2.15\tablenotemark{c} &      &$\gtrsim$32.7 & Probable (SBF, color)\\
\enddata
\tablenotetext{~}{Note. Units of right ascension are hours, minutes,
and seconds, and units of declination are degrees, arcminutes, and
arcseconds.}  \tablenotetext{a}{~Identifications from \citet{cote03}
for VCC1226 (M49), or from \citet{hanes01} for VCC1316 (M87).}
\tablenotetext{b}{~Average radial velocity from \citet{cote03} for
VCC1226, and from \citet{hanes01} for VCC1316. The velocity for S1063
is from \citet{huchra87}.}  \tablenotetext{c}{~c = 2.15 is the maximum
allowed value of the concentration index in the King models fits to
the surface brightness profiles.}
\end{deluxetable}

\clearpage

\clearpage

\clearpage

\begin{figure}
\plotone{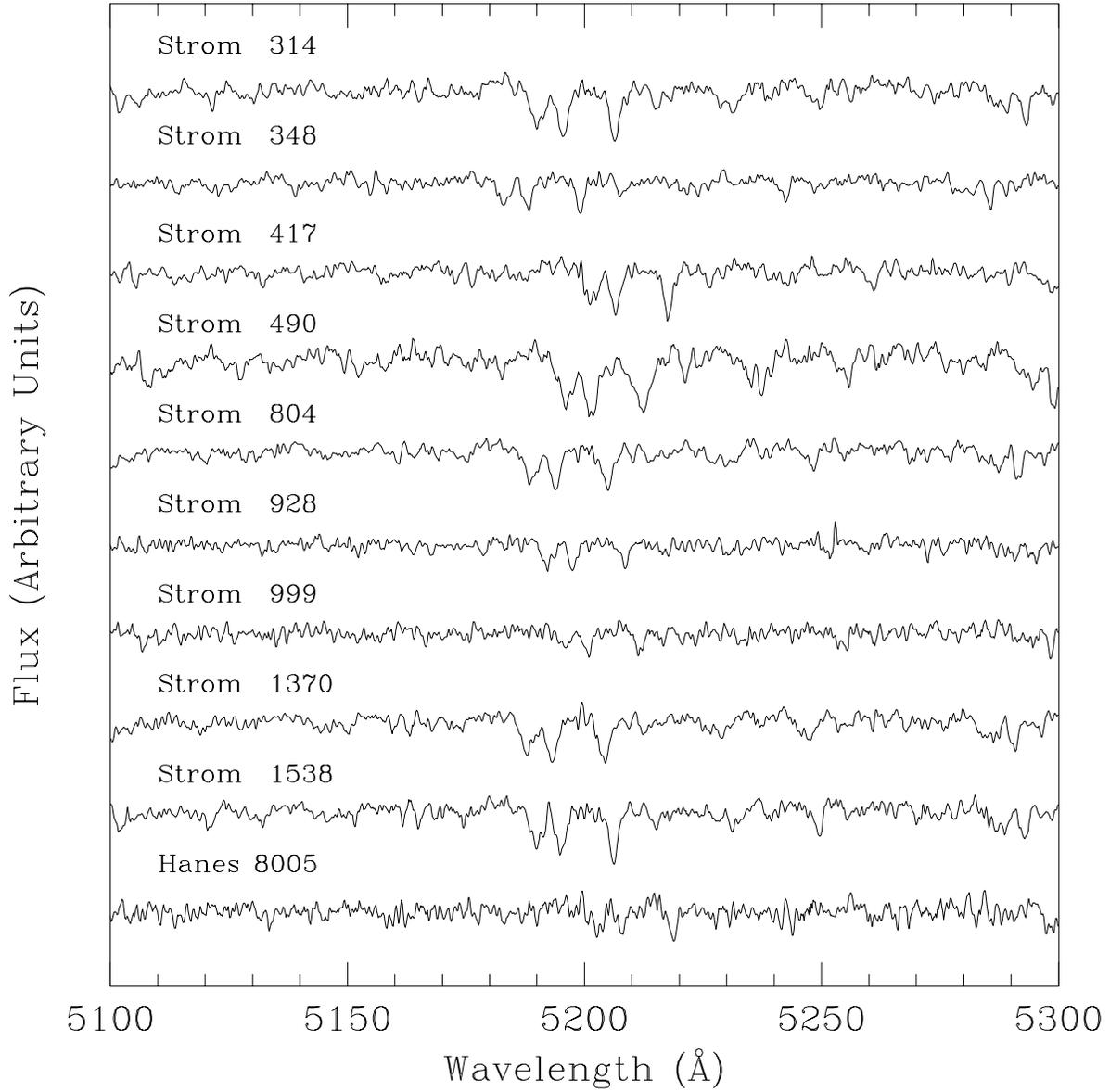}
\caption{Final ESI spectra showing a portion of order \#12 for each
program object.  The spectra have been smoothed with a boxcar having a
width of 3 pixels.
\label{fig03}}
\end{figure}

\clearpage

\begin{figure}
\plotone{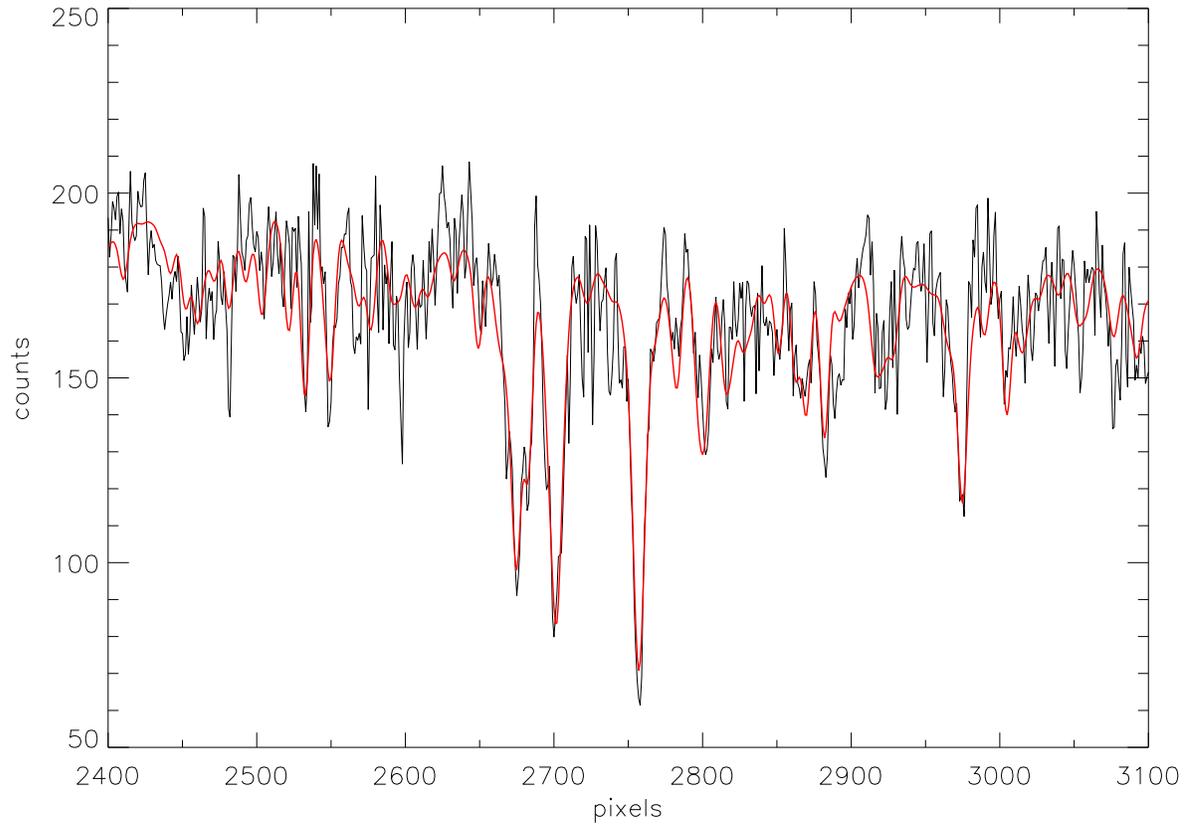}
\caption{pPXF plots in the approximate wavelength range 5100 to
5300{\AA}, showing the spectrum for S1538 in black, with the broadened
and fitted spectrum for the template (HD154417, an F8.5 IV-V star)
overlaid in red.
\label{fig04}}
\end{figure}

\clearpage

\begin{figure}
\plotone{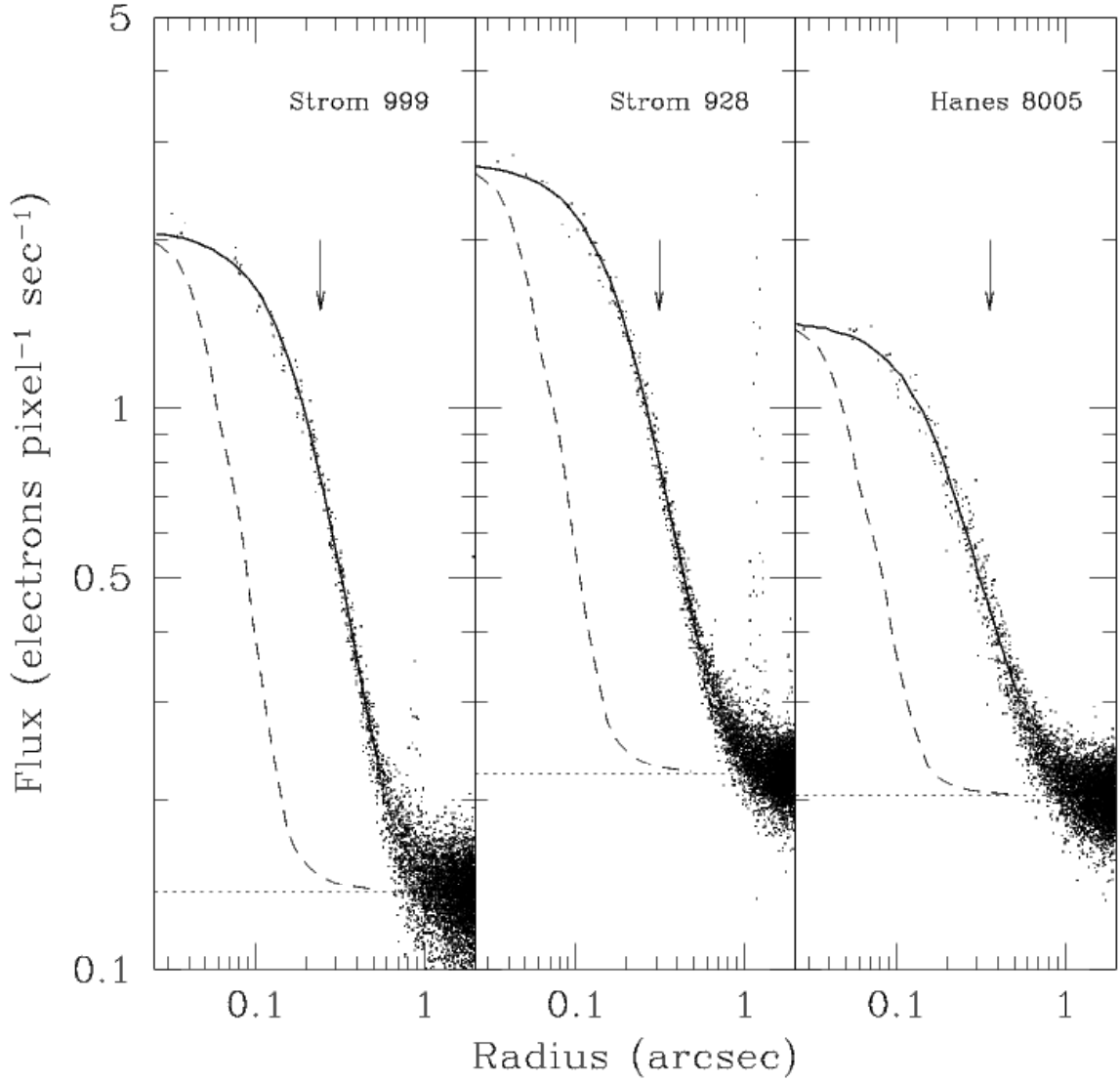}
\caption{Radial profiles for S999, S928 and H8005 measured in the
final F475W image. The adopted background is indicated by the dotted
line. The solid curve shows the best-fit, PSF-convolved King
model. The dashed curve shows the mean PSF for M87. The arrow
indicates the fitted half-light radius.
\label{fig05}}
\end{figure}

\clearpage

\begin{figure}
\plotone{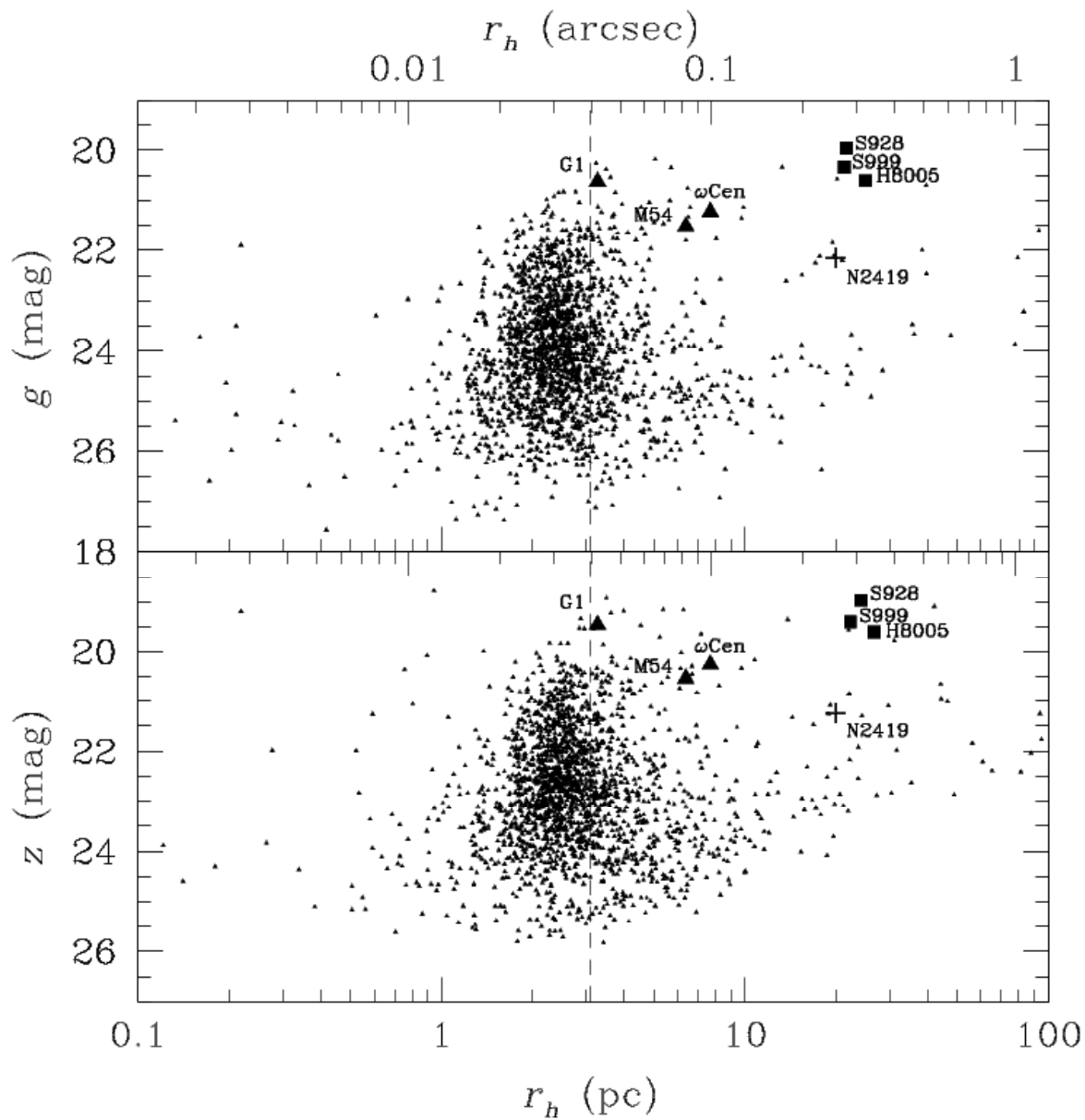}
\caption{Half-light radius, $r_h$, versus magnitude for $\approx$ 2000
sources detected in our M87 field. Results for $g$ and $z$ are
presented in the upper and lower panels, respectively.  The three
DGTOs with ESI spectra are shown by the large squares.  Three massive
globular clusters in the Local Group --- $\omega$ Cen, G1 and M54 ---
are shown as they would appear in our survey if located at the
distance of M87. The luminous and spatially extended Galactic globular
cluster NGC~2419 is shown by the cross. The vertical line in each
panel indicates the median half-light radius of globular clusters in
the Milky Way.
\label{fig06}}
\end{figure}

\clearpage

\begin{figure}
\epsscale{.8} \plotone{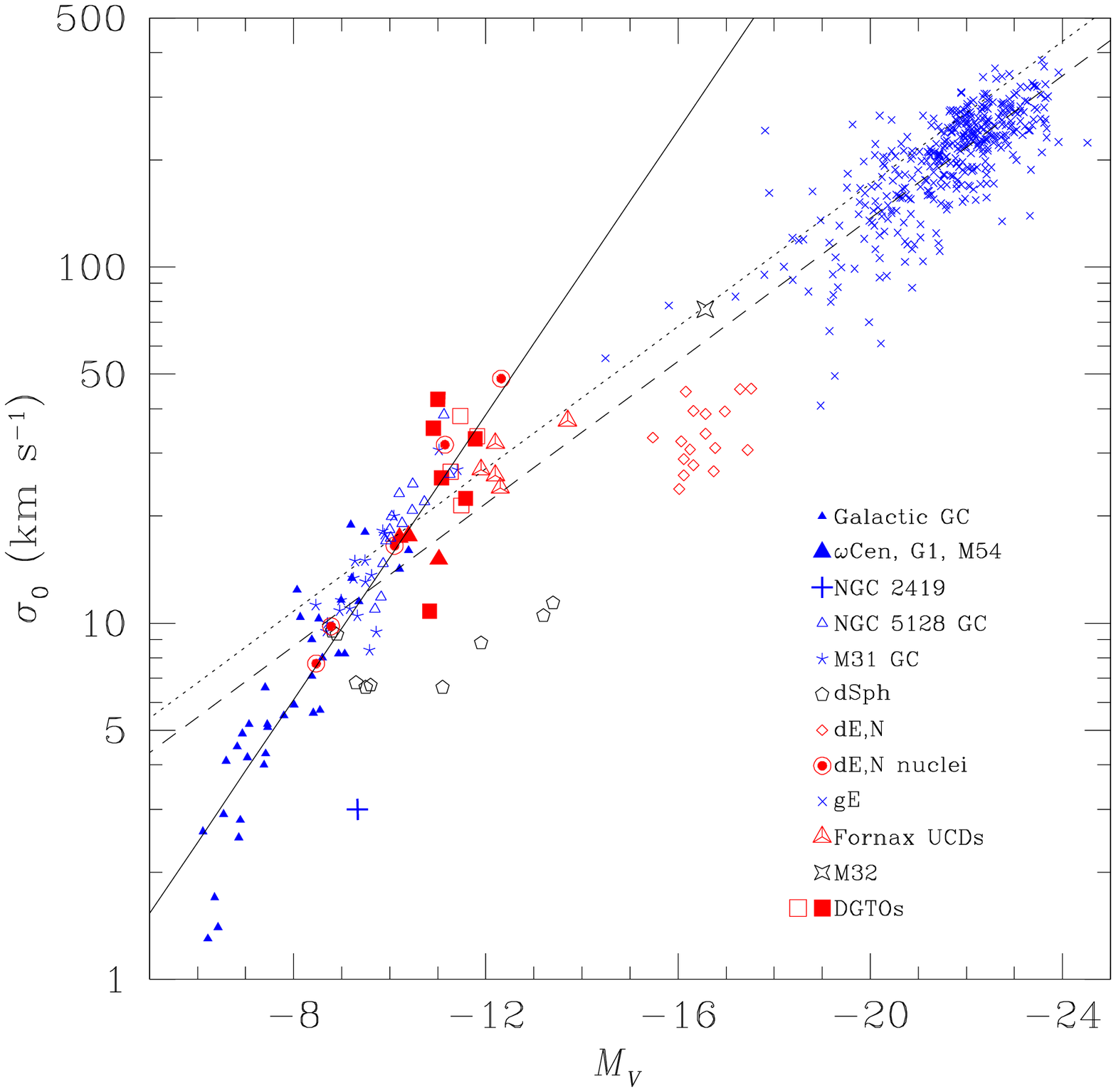}
\caption{Central velocity dispersion versus absolute visual magnitude
for hot stellar systems.  The plotted symbols show the location of
Galactic globular clusters \citep{mclaugh05} , M31 globular clusters
(a collation of data from \citealt{mclaugh05}, in preparation),
NGC~5128 globular clusters \citep{martin04,harris02}, Local Group dSph
galaxies \citep{mateo98}, Virgo dE,N galaxies and their nuclei
\citep{geha02}, the Fornax UCDs from \citet{drink03}, the compact
Local Group elliptical galaxy M32 \citep{mateo98,marel98,graham02} and
the giant elliptical galaxies of \citet{faber89}. $\omega$ Cen, G1 and
M54 are indicated by the large triangles, and NGC~2419 by the
cross. Our ten DGTOs are plotted as the large squares; {\it filled
squares} show the six objects with {\it HST} imaging, {\it open
squares} the four without.  For the latter objects, we plot their
observed velocity dispersions rather than the central values.  The
dashed line indicates our fit of the Faber-Jackson relation, $\sigma_0
\propto L^{0.25}$, to the giant ellipticals; the relation of
\citet{drink03} is shown by the dotted line.  The solid line indicates
the least-squares fit for Galactic globular clusters.
\label{fig07}}
\end{figure}

\clearpage

\begin{figure}
\plotone{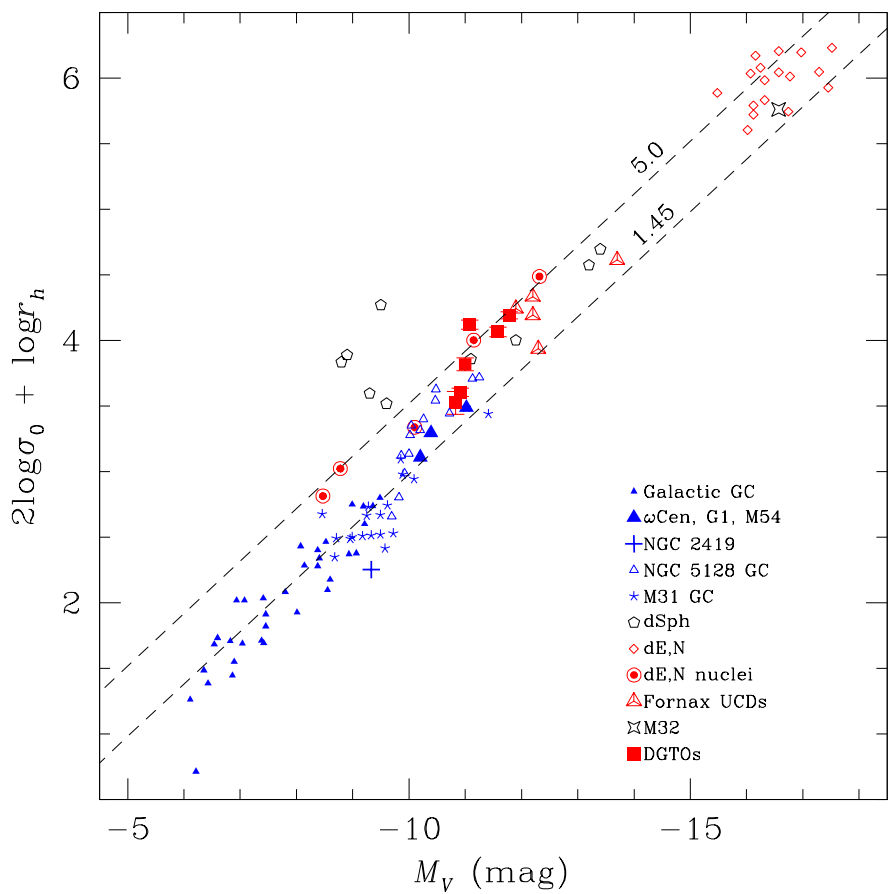}
\caption{A representation of the Virial Theorem for hot stellar
systems.  The symbols are the same as in Figure~\ref{fig07}.  The
lower dashed line shows the virial theorem for a constant
mass-to-light ratio of $\cal M/L_V$ = 1.45, the mean for Galactic
globular clusters \citep{mclaugh00}. A mass-to-light ratio of $\cal
M/L_V$ = 5, which is more typical of those measured for our program
objects, is indicated by the upper dashed line.
\label{fig08}}
\end{figure}

\clearpage

\begin{figure}
\plotone{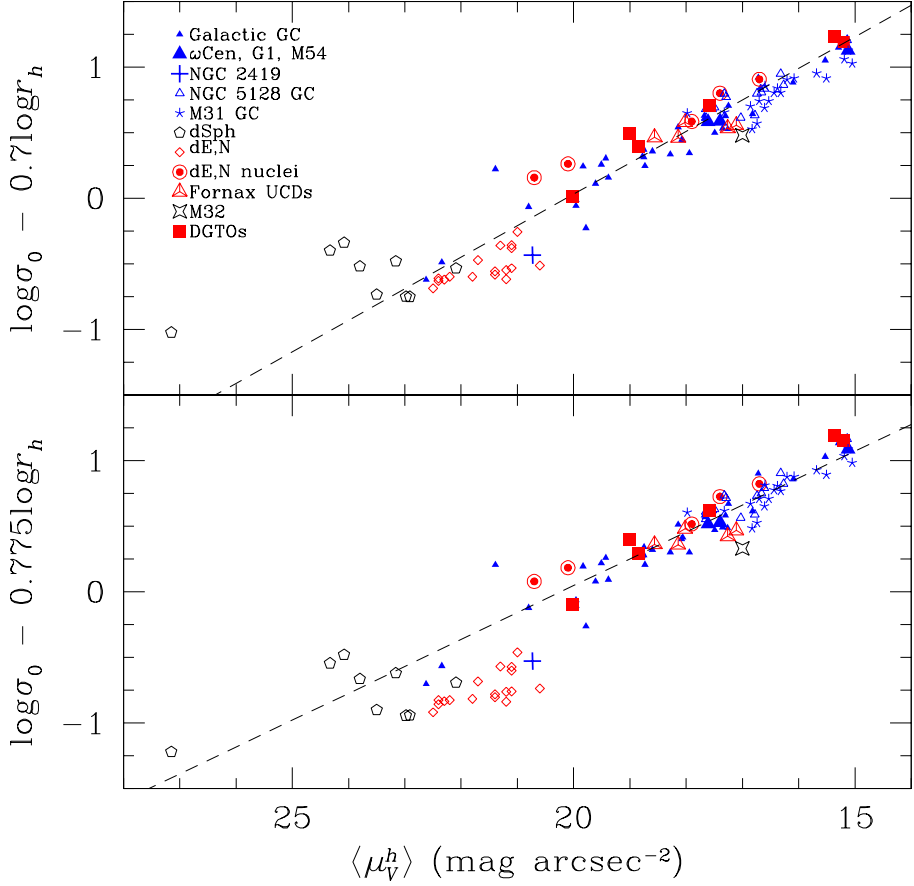}
\caption{{\it (Upper Panel)} The Fundamental Plane for globular
clusters in terms of half-light parameters, $r_h$ and
$\langle{\mu_V^h}\rangle$, and central velocity dispersion, $\sigma_0$
(from \citealt{djor95}).  The symbols are the same as in
Figure~\ref{fig07}. The dashed line shows the fitted relation for
Galactic globular clusters.  {\it (Lower Panel)} Alternative
representation of the globular cluster Fundamental Plane, following
\citet{mclaugh00}. The dashed line shows the relation between globular
cluster binding energy and luminosity, $E_b \propto L^{2.05}$.
\label{fig09}}
\end{figure}

\clearpage

\begin{figure}
\plotone{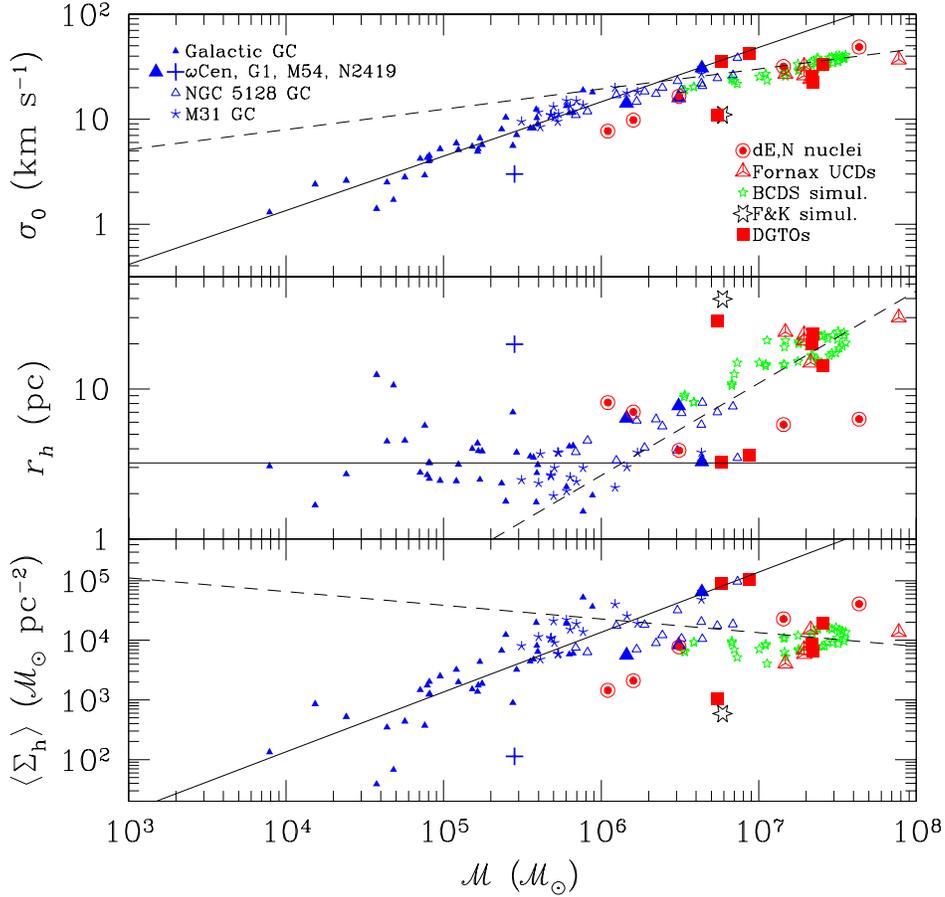}
\caption{Scaling relations for low-mass, hot stellar systems: i.e.,
central velocity dispersion, half-light radius, and mass surface
density averaged within the half-light radius, plotted against total
mass.  The symbols are the same as in Figure~\ref{fig07}, although we
now include the simulations of \citet{bekki03} and \citet{fellhauer02}
as small and large stars, respectively.  The dashed line in each panel
shows the extrapolation of the scaling relation for luminous
elliptical galaxies (see text for details).  The solid lines in the
upper and lower panels show the least-square fits to globular clusters
in the Galaxy; in the middle panel, the solid line shows the median
half-light radius of $r_h$ = 3.2 pc for Galactic globular clusters.
\label{fig10}}
\end{figure}

\clearpage

\begin{figure}
\plotone{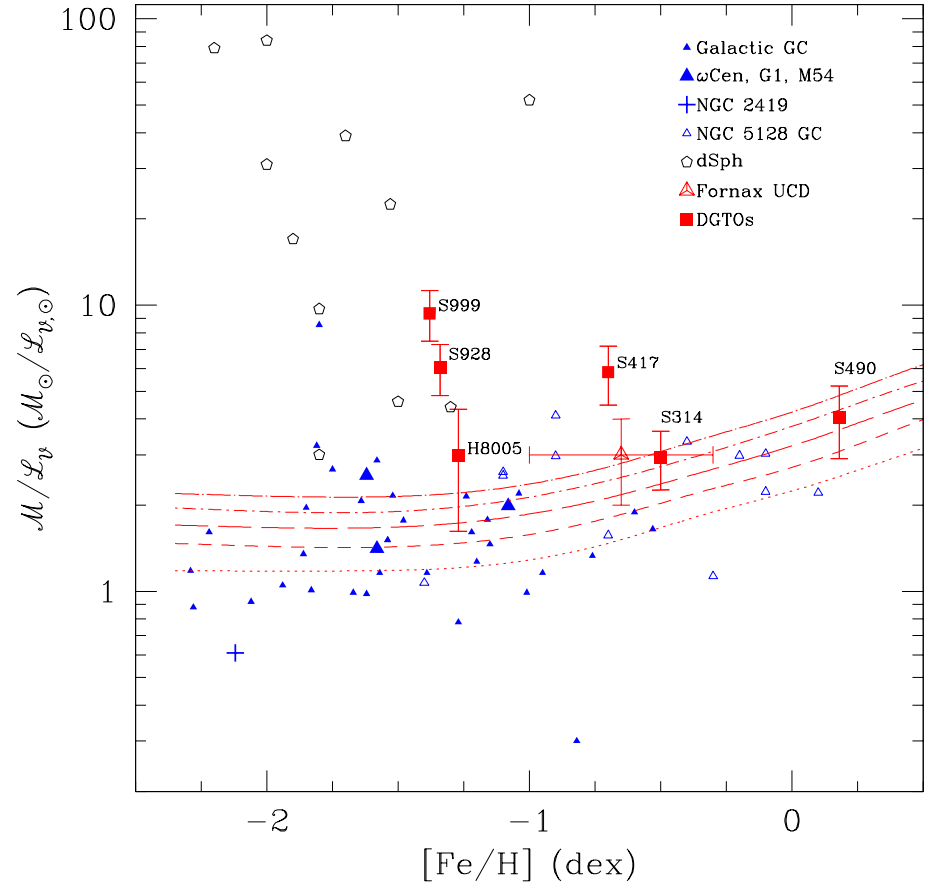}
\caption{Mass-to-light ratio versus metallicity for hot stellar
systems.  The symbols are the same as in Figure~\ref{fig07}.  From
bottom to top, the five curves show the theoretical predictions of the
population synthesis models of \citet{bruzual03} for ages of 7, 9, 11,
13 and 15 Gyr. These models assume a \citet{chabrier03} initial mass
function.
\label{fig11}}
\end{figure}

\clearpage

\begin{figure}
\plotone{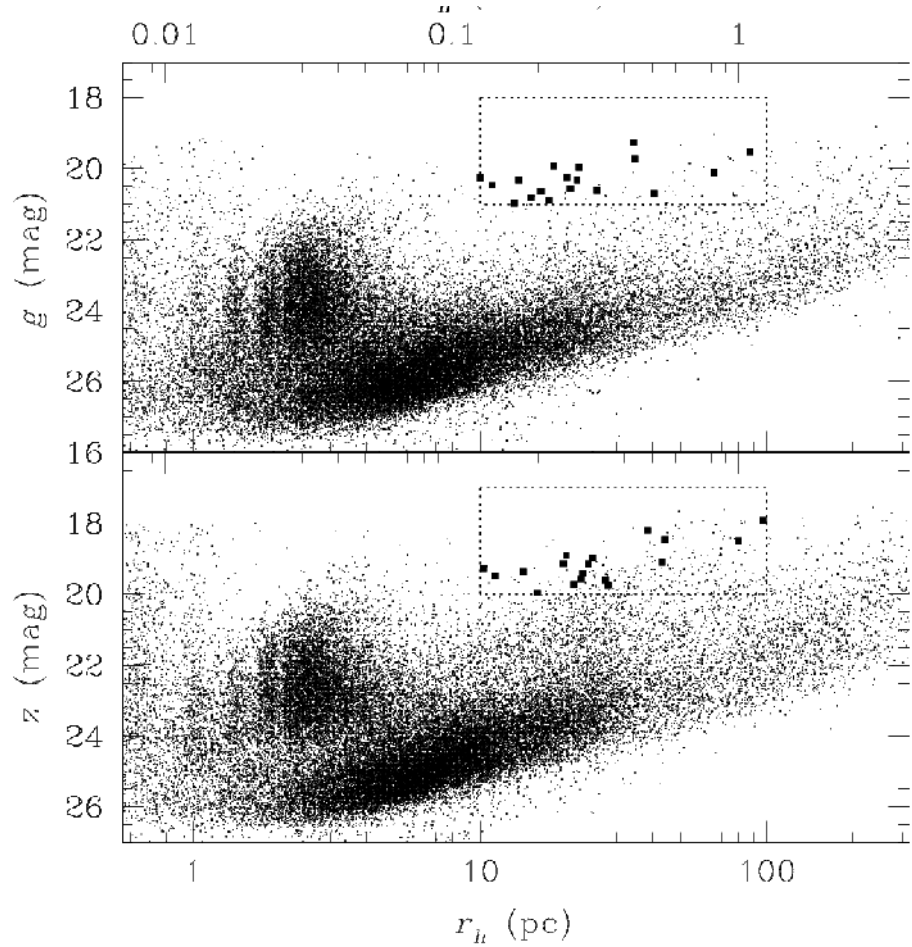}
\caption{Selection of DGTO candidates from the ACS Virgo Cluster
Survey based on magnitude, half-light radius and color. Points show
all cataloged objects in the 100 Virgo galaxy fields.  The large
squares enclosed by the dashed rectangular box are the 18 DGTO
candidates in Table 9.
\label{fig12}}
\end{figure}

\begin{figure}
\plotone{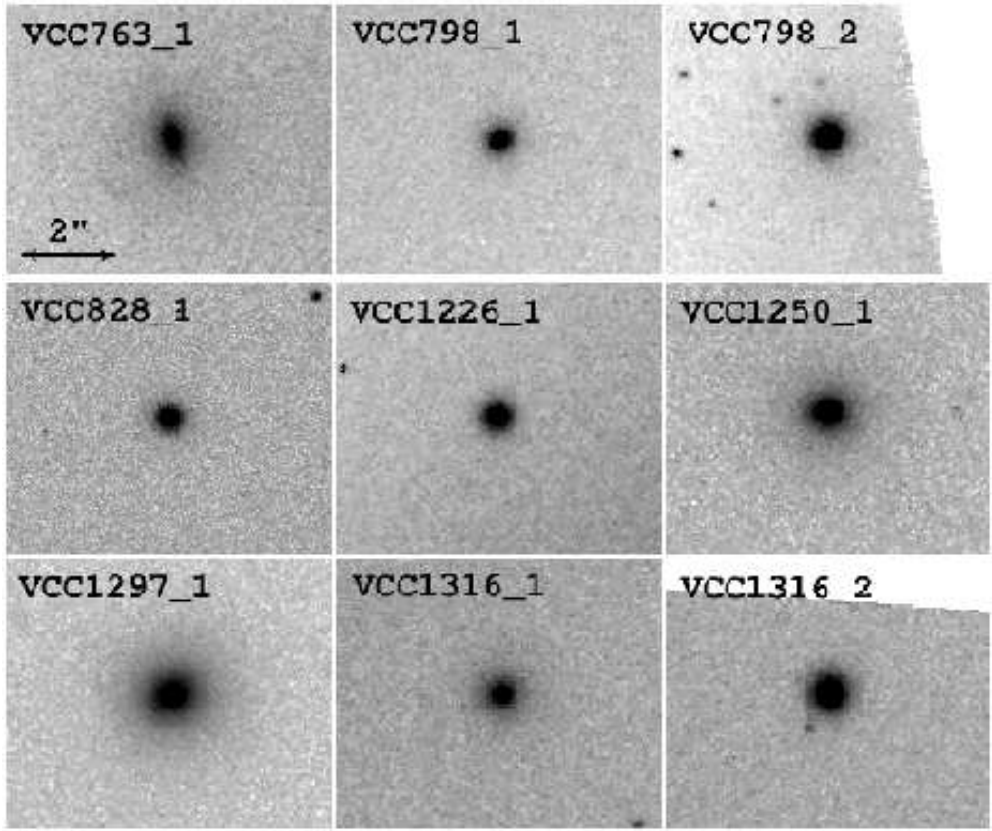}
\caption{Magnified F475W images for nine DGTO candidates from
Table~9. The line in the first panel has a length of 2\arcsec. These
objects were selected from the ACS Virgo Cluster Survey object
database for 100 program galaxies, using the magnitude, color, size
criteria described in \S\ref{sec:candidates}. North is up and East is
to the left in each panel.
\label{fig13}}
\end{figure}

\begin{figure}
\plotone{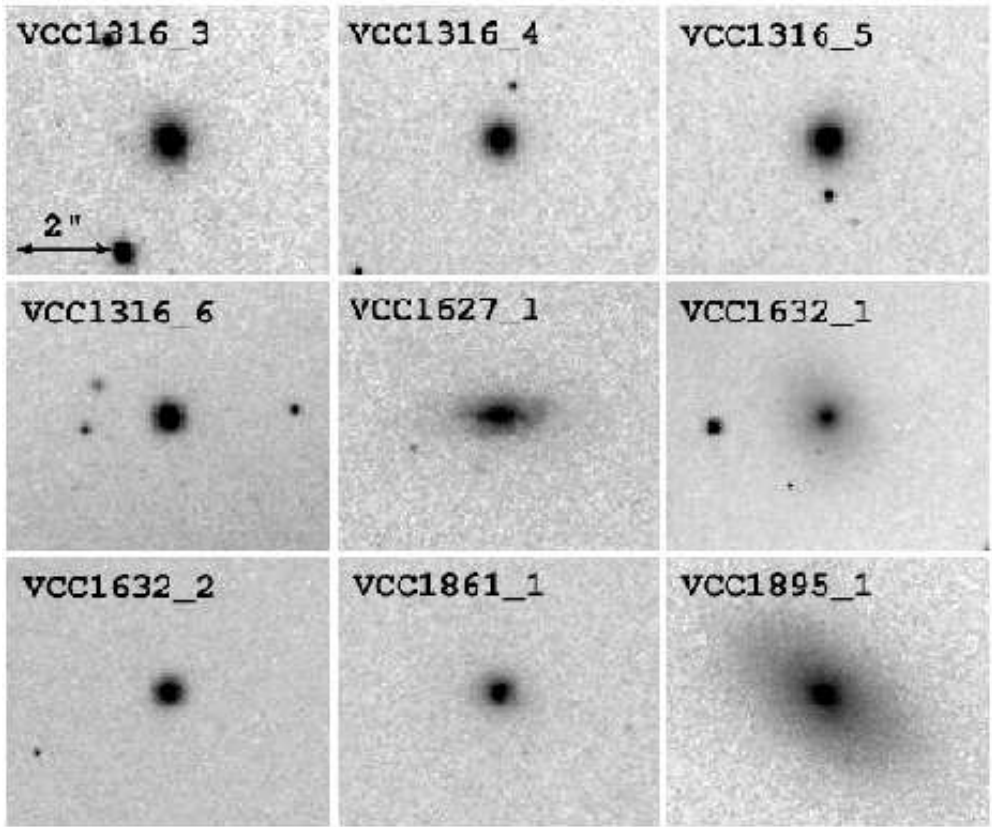}
\caption{Magnified images in the F475W filter for nine additional DGTO
candidates from Table~9. See Figure~\ref{fig13} for details.
\label{fig14}}
\end{figure}

\begin{figure}
\plotone{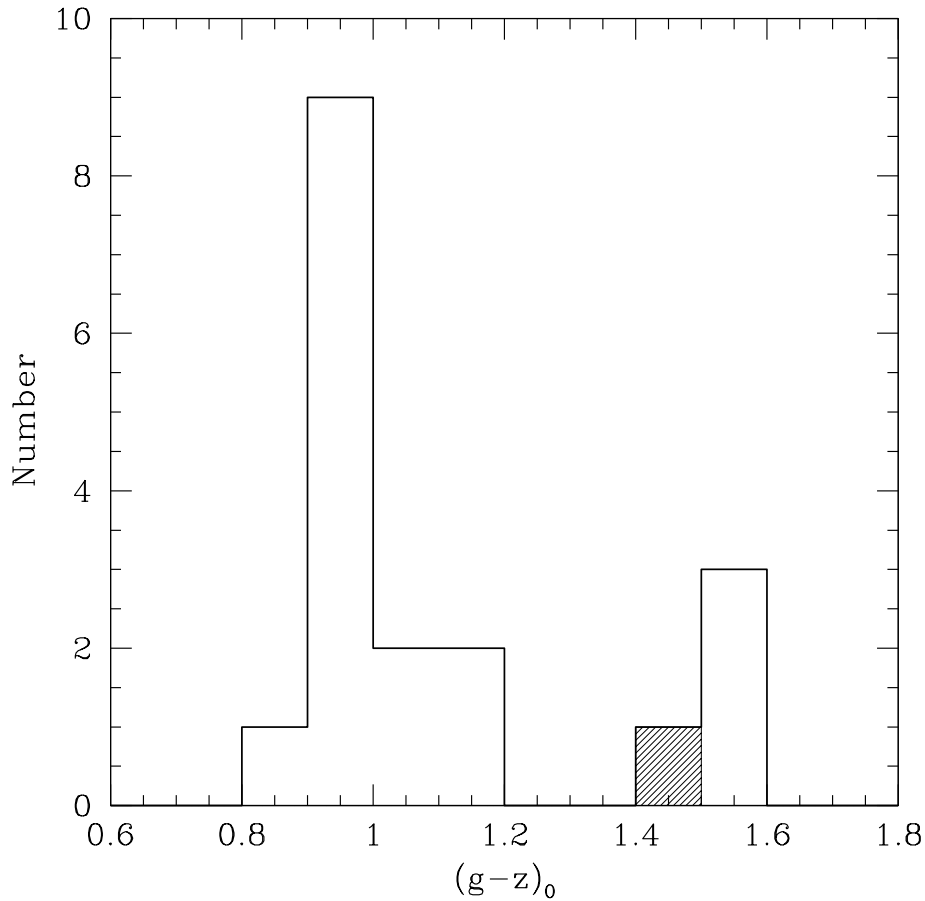}
\caption{Histogram of dereddened ($g-z$)$_0$ color for 18 DGTO
candidates selected from imaging in 100 fields in the Virgo Cluster
(open histogram). We identify two main populations among these
objects: probable or certain DGTOs with $0.88 \lesssim (g-z)_0
\lesssim 1.18$, and a redder population of probable background
galaxies. The shaded histogram shows the lone object in 17 control
fields from the ACS Virgo Cluster Survey which met our selection
criteria for DGTOs. Based on its red color, this object is likely to
be a background elliptical galaxy.
\label{fig15}}
\end{figure}

\begin{figure}
\plotone{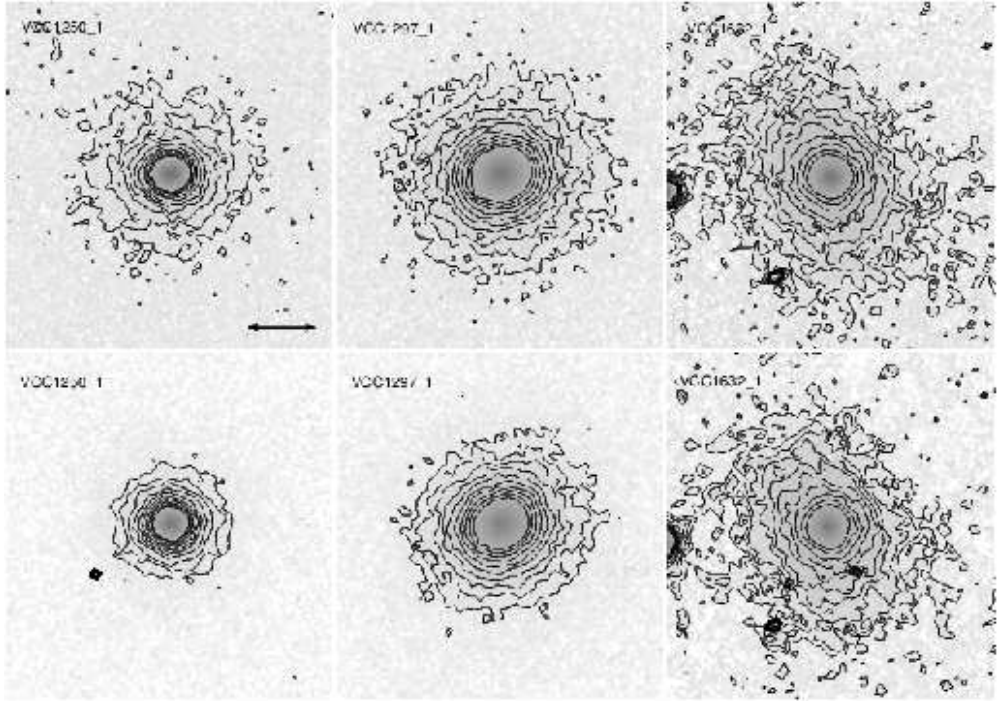}
\caption{Magnified views of VCC1250\_1, VCC1297\_1 and VCC1632\_1 in
the F475W and F850LP filters (upper and lower panels,
respectively). The light of the nearby galaxy has been subtracted in
each case, and contours have been overplotted to highlight the diffuse
envelopes.  North is up and East is to the left.
\label{fig16}}
\end{figure}

\begin{figure}
\plotone{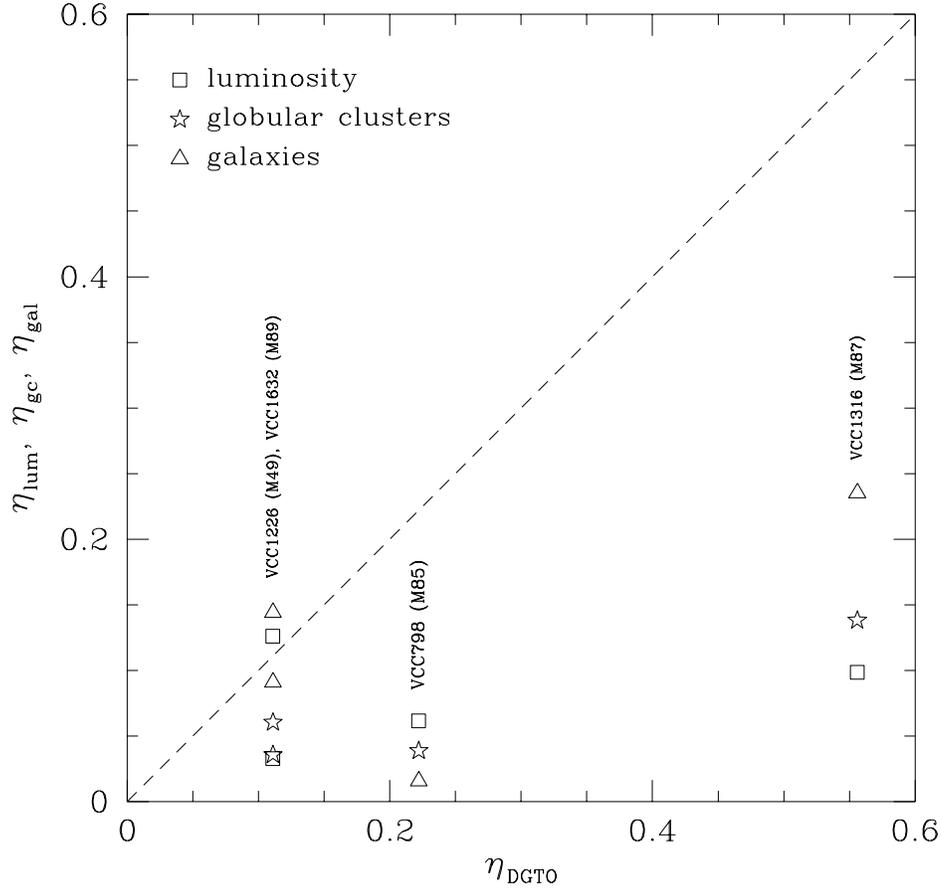}
\caption{Relative contribution of galaxies with DGTOs classified as
``certain" members of Virgo to the total DGTOs population, $\eta_{\rm
DGTO}$, plotted against: (1) the fraction of the blue luminosity
contained within each galaxy relative to the ACS Virgo Cluster Survey
sample ($\eta_{\rm lum}$; squares); (2) the fraction of high
probability globular cluster candidates in each galaxy relative to the
full survey ($\eta_{\rm gc}$; stars); and (3) the fraction of the 889
early-type VCC member galaxies with $B_T \ge 14$ which are found
within 1\fdg5~of each galaxy ($\eta_{\rm gal}$; triangles).  M87
(VCC1316) is more abundant in DGTOs than would be expected on the
basis of its luminosity, the size of its globular cluster system, or
the local galaxy density.
\label{fig17}}
\end{figure}

\end{document}